\documentclass[letterpaper,twocolumn,10pt]{article}
\usepackage{usenix}
\pdfoutput=1

\usepackage[hyphens]{url}
\usepackage{array,graphicx}
\usepackage{color}
\usepackage{caption}
\usepackage{subfigure}
\usepackage{multirow}
\usepackage{tabu}
\usepackage{paralist}
\usepackage{adjustbox}
\usepackage{booktabs}
\usepackage{amsmath}
\usepackage{hyperref}

\usepackage[compact]{titlesec}

\usepackage[font=small,labelfont=bf]{caption}

\usepackage{enumitem}

\newcommand*\rot{\rotatebox{90}}

\newcolumntype{C}[1]{>{\centering\let\newline\\\arraybackslash\hspace{0pt}}m{#1}}

\begin{document}

\title{Sound-Proof: Usable Two-Factor Authentication Based on Ambient Sound}

\author{
{\rm Nikolaos Karapanos, Claudio Marforio, Claudio Soriente and Srdjan \v{C}apkun} \\
Institute of Information Security \\ ETH Zurich \\
{\{firstname.lastname\}@inf.ethz.ch}}

\maketitle

\subsection*{Abstract}
Two-factor authentication protects online accounts even if passwords are leaked. Most users, however, prefer password-only authentication. One reason why two-factor authentication is so unpopular is the extra steps that the user must complete in order to log in. Currently deployed two-factor authentication mechanisms require the user to interact with his phone to, for example, copy a verification code to the browser. Two-factor authentication schemes that eliminate user-phone interaction exist, but require additional software to be deployed.

In this paper we propose Sound-Proof, a usable and deployable two-factor authentication mechanism. Sound-Proof does not require interaction between the user and his phone. In Sound-Proof the second authentication factor is the proximity of the user's phone to the device being used to log in. The proximity of the two devices is verified by comparing the ambient noise recorded by their microphones. Audio recording and comparison are transparent to the user, so that the user experience is similar to the one of password-only authentication. Sound-Proof can be easily deployed as it works with current phones and major browsers without plugins. We build a prototype for both Android and iOS. We provide empirical evidence that ambient noise is a robust discriminant to determine the proximity of two devices both indoors and outdoors, and even if the phone is in a pocket or purse. We conduct a user study designed to compare the perceived usability of Sound-Proof with Google 2-Step Verification. Participants ranked Sound-Proof as more usable and the majority would be willing to use Sound-Proof even for scenarios in which two-factor authentication is optional.



\section{Introduction}

Software tokens on modern phones are replacing dedicated hardware tokens in two-factor authentication (2FA) mechanisms. Using a software token, in place of a hardware one, improves deployability and usability of 2FA. For service providers, 2FA based on software tokens results in a substantial reduction of manufacturing and shipping costs.
From the user's perspective, there is no extra hardware to carry around and phones can accommodate software tokens from multiple service providers.

Despite the improvements introduced by software tokens, most users still prefer password-only authentication for services where 2FA is not mandatory~\cite{petsas15eurosec,imperi13umsurvey}.
This is probably due to the extra burden that 2FA causes to the user~\cite{gunson11cs,weir10int}, since it typically requires the user to interact with his phone.

Recent work~\cite{czeskis12ccs, shirvanian14} improves the usability of 2FA by eliminating the user-phone interaction. However, those proposals are not yet deployable as their requirements are not met by today's phones, computers or browsers.

In this paper, we focus on both the usability and deployability aspect of 2FA solutions.
We propose Sound-Proof, a two-factor authentication mechanism that is transparent to the user and can be used with current phones and with major browsers without any plugin.
In Sound-Proof the second authentication factor is the proximity of the user's phone to the computer being used to log in.
When the user logs in, the two devices record the ambient noise via their microphones.
The phone compares the two recordings, determines if the computer is located in the same environment, and ultimately decides whether the login attempt is legitimate or fraudulent.

Sound-Proof does not require the user to interact with his phone. The overall user experience is, therefore, close to password-only authentication.
Sound-Proof works even if the phone is in the user's pocket or purse, and both indoors and outdoors. Sound-Proof can be easily deployed since it is compatible with current phones, computers and browsers. In particular, it works with any HTML5-compliant browser that implements the WebRTC API~\cite{webrtc}, which is currently being standardized by the W3C~\cite{webrtcw3c}.
Chrome, Firefox and Opera already support WebRTC, Internet Explorer plans to support it~\cite{iewebrtc}, and we anticipate that other browsers will adopt it soon.

Similar to other approaches that do not require user-phone interaction nor a secure channel between the phone and the computer (e.g.,~\cite{czeskis12ccs}),
Sound-Proof is not designed to protect against targeted attacks where the attacker is co-located with the victim and has the victim's login credentials.
Our design choice favors usability and deployability over security and we argue that this can edge for larger user adoption.

We have implemented a prototype of Sound-Proof for both Android and iOS.
Sound-Proof adds, on average, less than 5 seconds to a password-only login operation.
This time is substantially shorter than the time overhead of 2FA mechanisms based on verification codes (roughly 25 seconds~\cite{weir09compsec}).
We also report on a user study we conducted which shows that users prefer Sound-Proof over Google 2-Step Verification~\cite{google2step}.

In summary, we make the following contributions:

\begin{itemize}
\item   We propose Sound-Proof, a novel 2FA mechanism that does not require user-phone interaction and is easily deployable.
        The second authentication factor is the proximity of the user's phone to the computer from which he is logging in.
        Proximity of the two devices is verified by comparing the ambient audio recorded via their microphones. Recording and comparison are transparent to the user.

\item   We implement a prototype of our solution for both Android and iOS. We use the prototype to evaluate the effectiveness of Sound-Proof in a number of different settings.
        We show that Sound-Proof works even if the phone is in the user's pocket or purse and that it fares well both indoors and outdoors.

\item   We conducted a user study to compare the perceived usability of Sound-Proof and Google 2-Step Verification.
        Participants ranked the usability of Sound-Proof higher than the one of Google 2-Step Verification, with a statistically significant difference.
        More importantly, we found that most participants would use Sound-Proof even if 2FA were optional.
\end{itemize}

The rest of the paper is organized as follows.
Section~\ref{sec:goals} details our assumptions and goals while Section~\ref{sec:alternative} reviews alternative approaches and discusses why they do not fulfill our objectives.
Section~\ref{sec:background} provides an overview on audio similarity techniques.
We present Sound-Proof in Section~\ref{sec:architecture} and its prototype implementation in Section~\ref{sec:implementation}.
Section~\ref{sec:evaluation} evaluates Sound-Proof, while Section~\ref{sec:userstudy} reports on our user study.
We discuss limitations and ways to further improve Sound-Proof in Section~\ref{sec:discussion}.
Section~\ref{sec:relatedwork} reviews related work and Section~\ref{sec:conclusion} concludes the paper.

\section{Assumptions and Goals}
\label{sec:goals}

\noindent\textbf{System Model.}
We assume the general settings of browser-based web authentication.
The user has a username and a password to authenticate to a web server.
The server implements a 2FA mechanism that uses software tokens on phones.

The user points his browser to the server's webpage and enters his username and password.
The server verifies the validity of the password and challenges the user to prove possession of the second authentication factor.

\noindent\textbf{Threat Model.}
We assume a remote adversary who has obtained the victim's username and password via phishing, leakage of a password database, or via other means.
His goal is to authenticate to the server on behalf of the user.
In particular, the adversary visits  the server's webpage and enters the username and password of the victim.
The attack is successful if the adversary convinces the server that he also holds the second authentication factor of the victim.

We further assume that the adversary cannot compromise the victim's phone.
If the adversary gains control of the platform where the software token runs, then the security of any 2FA scheme reduces to the security of password-only authentication.
Also, the adversary cannot compromise the victim's computer.
The compromise of the computer allows the adversary to mount a Man-In-The-Browser attack~\cite{owasp_mitb} and hijack the victim's session with the server, therefore defeating any 2FA mechanism.

We do not address targeted attacks where the adversary is co-located with the victim.
2FA mechanisms that do not require the user to interact with his phone cannot protect against targeted, co-located attacks.
For example, if 2FA uses unauthenticated short-range communication~\cite{czeskis12ccs}, a co-located attacker can connect to the victim's phone and prove possession of the second authentication factor to the server.
We argue that targeted, co-located attacks are less common than non-selective, remote attacks.
Furthermore, any 2FA mechanism may not warrant protection against powerful attackers.
For example, if 2FA uses verification codes, a determined attacker may gain physical access to the phone or read the code from a distance~\cite{backes09sp,backes08sp,raguram11ccs}.


We do not consider Man-In-The-Middle adversaries. Client authentication is not sufficient to defeat MITM attacks in the context of web applications~\cite{karapanos14usenix}.
We also do not address active phishing attacks where the attacker lures the user into visiting a phishing website and relays the stolen credentials to the legitimate website in real-time. Such attacks can be thwarted by having the phone detect the phishing domain~\cite{czeskis12ccs,parno06fc}. This requires short-range communication between the phone and the browser. However, seamless short-range communication between the phone and the browser is currently not possible.

\noindent\textbf{Design Goals.}

\begin{itemize}

\item   \emph{Usability.}
        Users should authenticate using only their username and password as in password-only authentication.
        In particular, users should not be asked to interact with their phone --- not even to pick up the phone or take it out of a pocket or purse.
\item   \emph{Deployability.}
        The 2FA mechanism should work with common smartphones, computers and browsers.
        It should not require additional software on the computer or the installation of browser plugins.
        A plugin-based solution limits the usability of the system because (\emph{i}) a different plugin may be required for each server,
        and (\emph{ii}) the user must install the plugin every time he logs in from a computer for the first time.
        The mechanism should also work on a wide range of smartphones.
        We therefore discard the use of special hardware on the phone like NFC chips or biometric sensors.
\end{itemize}

\section{Alternative Approaches}
\label{sec:alternative}

In this section we discuss traditional 2FA mechanisms, as well as 2FA proposals which minimize the user-phone interaction.
For each solution we argue why it fails to meet our usability and deployability goals.


\subsection{Traditional 2FA}
\label{sec:traditional}

\noindent\textbf{Hardware Tokens.}
Hardware tokens range from the RSA SecurID~\cite{rsa} to recent dongles~\cite{yubico} that comply with the FIDO U2F~\cite{fido} standard for universal 2FA.
Such solutions require the user to carry and interact with the token and may be expensive to deploy because the service provider must ship one token per customer.

\noindent\textbf{Software Tokens.}
Google 2-Step Verification~\cite{google2step} is an example of 2FA based on verification codes, that uses software tokens on phones. The verification code is retrieved either from an application running on the phone or via SMS. Such mechanisms require the user to copy the verification code from the phone to the browser.

Duo Push~\cite{duosecurity} and Encap Security~\cite{encap} prompt the user with a push message on his phone with information on the current login attempt.
Both solutions still require the user to interact with his phone to authorize the login.


\subsection{Reduced-Interaction 2FA}
\label{sec:minimal}

\noindent\textbf{Short-range Radio Communication.}
PhoneAuth~\cite{czeskis12ccs} is a 2FA proposal that leverages unpaired Bluetooth communication between the browser and the phone, in order to eliminate user-phone interaction.
The Bluetooth channel enables the server (through the browser) and the phone to engage in a challenge-response protocol which provides the second authentication factor.
Similarly,~\cite{parno06fc} and~\cite{shirvanian14} also leverage Bluetooth communication between the browser and the phone.

These schemes require the browser to expose a Bluetooth API that is currently not available on any browser.
A specification to expose a Bluetooth API in browsers has been proposed by the Web Bluetooth Community Group~\cite{w3cwebbluetooth}.
It is unclear whether the proposed API will support the unauthenticated RFCOMM or similar functionality which is required to enable seamless connectivity between the browser and the phone.
However, if the Bluetooth connection is unauthenticated, an adversary equipped with a powerful antenna may connect to the victim's phone from afar~\cite{aircable} and login on behalf of the user, despite 2FA.

Authy~\cite{authy} is another approach that allows for seamless 2FA using Bluetooth communication between the computer and the phone.
Authy, however, requires extra software on the computer.

As an alternative to Bluetooth, the browser and the phone can communicate over WiFi~\cite{shirvanian14}.
This approach only works when both devices are on the same network.
Shirvanian et al.,~\cite{shirvanian14} use extra software on the computer to virtualize the wireless interface and create a software access point (AP) with which the phone needs to be associated. The user has to perform this setup procedure every time he uses a new computer to log in.
Their solution also requires a phone application listening for incoming connections in the background, which is currently not possible on iOS.

Finally, the browser and the phone can communicate over NFC. NFC hardware is not commonly found in commodity computers, and current browsers do not expose APIs to access NFC. 
Furthermore, a solution based on NFC would not completely remove user-phone interaction because the user would still need to hold his phone close to the computer.

We acknowledge that 2FA mechanisms that employ direct communication between the browser and the phone may provide additional security against remote attackers.
For example, the phone can detect if the user tries to login on a phishing website and block the attempt~\cite{czeskis12ccs,parno06fc}.
The scheme in~\cite{shirvanian14} further resists offline dictionary attacks against compromised hashed password databases.
Nevertheless, none of such solutions can be deployed for the reasons we discussed above.

\noindent\textbf{Near-ultrasound.}
SlickLogin~\cite{slicklogin} minimizes the user-phone interaction transferring the verification code from the computer to the phone using near-ultrasounds.
The idea is to use spectrum frequencies that are non-audible for the majority of the population but that can be reproduced by the speakers of commodity computers ($>18$kHz).
Using non-audible frequencies accommodates for scenarios where users may not want their devices to make audible noise.
Due to their size, the speakers of commodity computers can only produce highly directional near-ultrasound frequencies~\cite{russell98amjphys}.
Near-ultrasound signals also attenuate faster, when compared to sounds in the lower part of the spectrum ($<18$kHz)~\cite{arentz11ubicomp,hazas02ubicomp}.
With SlickLogin, the user must ensure that the speaker volume is at a sufficient level during login.
Also, login will fail if a headset is plugged into the laptop.
Finally, this approach may not work in scenarios where there is in-band noise (e.g., when listening to music or in cafes)~\cite{hazas02ubicomp}.
We also note that a solution based on near-ultrasounds may result unpleasant for young people and animals that are capable of hearing sounds above 18kHz~\cite{valiente14audiology}.

\noindent\textbf{Location Information.}
The server can check if the computer and the phone are co-located by comparing their GPS coordinates.
GPS sensors are available on all modern phones but are rare on commodity computers.
If the computer from which the user logs in has no GPS sensor, it can use the geolocation API exposed by some browsers~\cite{firefoxgeolocation}.
Nevertheless, information retrieved via the geolocation API may not be accurate, for example when the device is behind a VPN or it is connected to a large managed network (such as enterprise or university networks). 
Furthermore, geolocation information can be easily guessed by an adversary.
For example, assume the adversary knows the location of the victim's workplace and uses that location as the second authentication factor.
This attack is likely to succeed during working hours since the victim is presumably at his workplace.

\noindent\textbf{Other Sensors.}
A 2FA mechanism can combine the readings of multiple sensors that measure ambient characteristics, such as temperature, concentration of gases in the atmosphere, humidity, and altitude, as proposed in~\cite{shrestha14}. These combined sensor modalities can be used to verify the proximity between the computer through which the user is trying to login and his phone. However, today's computers and phones lack the hardware sensors that are required for such an approach to work.

%

\section{Background on Sound Similarity}
\label{sec:background}
The problem of determining the similarity of two audio samples is close to the problem of audio fingerprinting and automatic media retrieval~\cite{chandrasekhar11ismir}.
In media retrieval, a noisy recording is matched against a database of reference samples.
This is done by extracting a set of relevant features from the noisy recording and comparing them against the features of the reference samples.
The extracted features must be robust to, for example, background noise and attenuation.
Bark Frequency Cepstrum Coefficients~\cite{haitsma02}, wavelets~\cite{baluja08pr} or peak frequencies~\cite{Wang06cacm} have been proposed as robust features for automatic media retrieval.
Such techniques focus mostly on the frequency domain representation of the samples because they deal with time-misaligned samples.
In our scenario, we compare two quasi-aligned samples (the offset is less than 150ms) and we therefore can also extract relevant information from their time domain representations.

In order to consider both time domain and frequency domain information of the recordings, we use one-third octave band filtering and cross-correlation.

\noindent\textbf{One-third Octave Bands.}
Octave bands split the audible range of frequencies (roughly from $20$Hz to $20$kHz) in $11$ non-overlapping bands where the ratio
of the highest in-band frequency to the lowest in-band frequency is $2$ to $1$.
Each octave is represented by its center frequency, where the center frequency of a particular octave is twice the center frequency of the previous octave.
One-third octave bands split the first 10 octave bands in three and the last octave band in two, for a total of $32$ bands.
One-third octave bands are widely used in acoustics and their frequency ranges have been standardized~\cite{ansi}.
The center frequency of the lowest band is $16$Hz (covering from $14.1$Hz to $17.8$Hz) while the center frequency of the highest band is $20$kHz (covering from $17780$Hz to $22390$Hz).
In the following we denote with $B=[lb-hb]$ a set of contiguous one-third octave bands, from the band that has its central frequency at $lb$Hz,
to the band that has its central frequency at $hb$Hz.

Splitting a signal in one-third octave bands provides high frequency resolution information of the original signal, while keeping its time-domain representation.

\begin{figure*}[t]
	\centering
	\includegraphics[width=.8\textwidth]{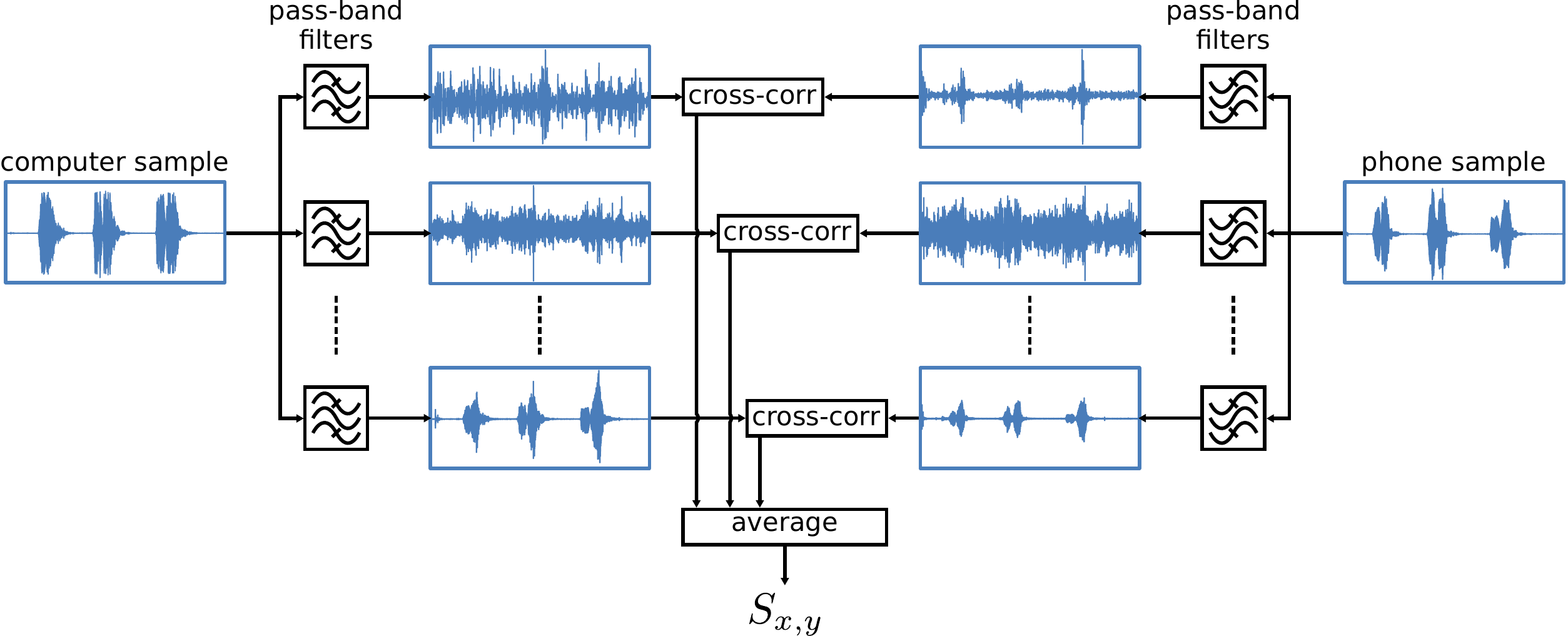}
	\caption{Block diagram of the function that computes the similarity score between two samples. The computation takes place on the phone. If $S_{x,y}>\tau_{C}$ and the average power of the samples is greater than $\tau_{dB}$, the phone judges the login attempt as legitimate.
 }
	\label{fig:similarity}
\end{figure*}

\noindent\textbf{Cross-correlation.}
Cross-correlation is a standard measure of similarity between two time series.
Let $x,\ y$ denote two signals represented as n-points discrete time series,\footnote{For simplicity we assume both series to have the same length.} the cross-correlation $c_{x,y}(l)$ measures their similarity as a function of the lag $l\in[0,n-1]$ applied to $y$:

$$
c_{x,y}(l)= \sum_{i=0}^{n-1}x(i)\cdot y(i-l)\\
$$

where $y(i)=0$ if $i<0$ or $i>n-1$.

To accommodate for different amplitudes of the two signals, the cross correlation can be normalized as:

$$
c'_{x,y}(l)=\frac{c_{x,y}(l)}{\sqrt{c_{x,x}(0)\cdot c_{y,y}(0)}}
$$

where $c_{x,x}(l)$ is known as auto-correlation.

The normalization maps $c'_{x,y}(l)$ in $[-1,1]$.
A value of $c'_{x,y}(l)=1$ indicates that at lag $l$, the two signals have the same shape even if their  amplitudes may be different;
a value of $c'_{x,y}(l)=-1$ indicates that the two signals have the same shape but opposite signs.
Finally, a value of $c'_{x,y}(l)=0$ shows that the two signals are uncorrelated.

If the actual lag between the two signals is unknown, we can discard the sign information and use the absolute value
of the maximum cross-correlation $\hat{c}_{x,y}(l)=\underset{l}{\operatorname{max}}(|c'_{x,y}(l)|)$
as a metric of similarity ($0\leq \hat{c}_{x,y}(l)\leq 1$).

The computation overhead of $c_{x,y}(l)$ can be decreased by leveraging the cross-correlation theorem
and computing $c_{x,y}(l)= F^{-1}(F(x)^* \cdot F(y))$, where $F()$ denotes the discrete Fourier transform and the asterisk denotes the complex conjugate.


\section{Sound-Proof Architecture}
\label{sec:architecture}

The second authentication factor of Sound-Proof is the proximity of the user's phone to the computer being used to log in.
The proximity of the two devices is determined by computing a similarity score between the ambient noise captured by their microphones.
For privacy reasons we do not upload cleartext audio samples to the server.
In our design, the computer encrypts its audio sample under the public key of the phone.
The phone receives the encrypted sample, decrypts it, and computes the similarity score between the received sample and the one recorded locally.
Finally, the phone tells the server whether the two devices are co-located or not. Note that the phone never uploads its recorded sample to the server.
Communication between the computer and the phone goes through the server.
We avoid short-range communication between the phone and the computer  (e.g., via Bluetooth) because it requires changes to the browser or the installation of a plugin.

\subsection{Similarity Score}
\label{sec:similarity}
Figure~\ref{fig:similarity} shows a block diagram of the function that computes the similarity score.
Each audio signal is input to a bank of pass-band filters to obtain $n$ signal components, one per each of the one-third octave bands that we take into account.
Let $x_i$ be the signal component for the $i$-th one-third octave band of signal $x$.
The similarity score is the average of the maximum cross-correlation over the pairs of signal components $x_i,\ y_i$:

$$S_{x,y} = \frac{1}{n}\sum_{i=1}^{i={n}}\hat{c}_{x_i,y_i}(l)$$

where $l$ is bounded between $0$ and $\ell_{max}$.

\begin{figure}
	\begin{center}
		\includegraphics[width=.9\columnwidth]{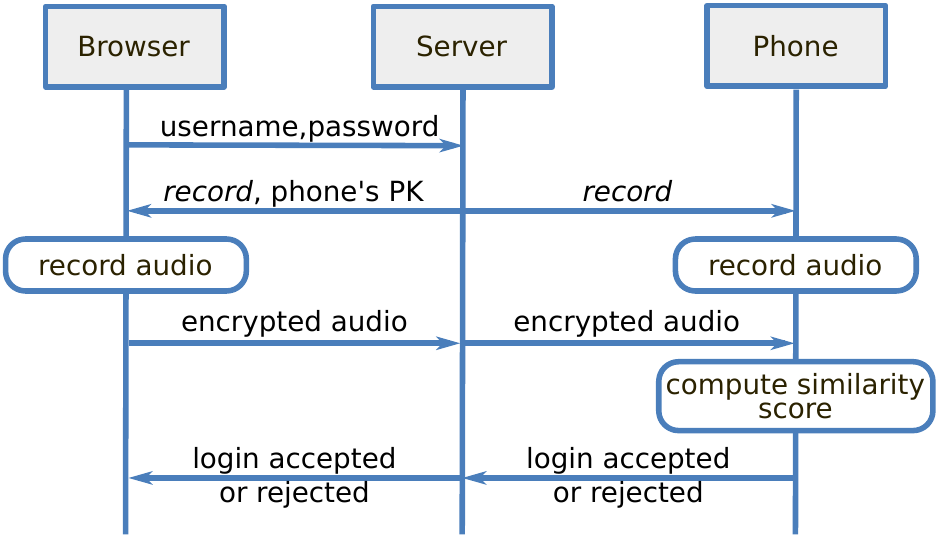}
		\caption{Sound-Proof authentication overview. At login, the phone and the computer record ambient noise with their microphones.
        The phone computes the similarity score between the two samples and returns the result to the server.}
		\label{fig:overview}
	\end{center}
\end{figure}

\subsection{Enrollment and Login}
Similar to other 2FA mechanisms based on software tokens,
Sound-Proof requires the user to install an application on his phone and to bind the application to his account on the server.
This one-time operation can be carried out using existing techniques to enroll software tokens, e.g.,~\cite{google2step}.
We assume that, at the end of the phone enrollment procedure, the server receives the unique public key of the application on the user's phone and binds that public key to the account of that user.

Figure~\ref{fig:overview} shows an overview of the login procedure.
The user points the browser to the URL of the server and enters his username and password.
The server retrieves the public key of the user's phone and sends it to the browser.
Both the browser and the phone start recording through their local microphones for $t$ seconds.
During recording, the two devices synchronize their clocks with the server.
When recording completes, each device adjusts the timestamp of its sample taking into account the clock difference with the server.
The browser encrypts the audio sample under the phone's public key and sends it to the phone, using the server as a proxy.
The phone decrypts the browser's sample and compares it against the one recorded locally. 
If the average power of both samples is above $\tau_{dB}$ and the similarity score is above $\tau_C$,
the phone concludes that it is co-located with the computer from which the user is logging in and informs the server that the login is legitimate.

The procedure is completely transparent to the user if the environment is sufficiently noisy.
In case the environment is quiet, Sound-Proof requires the user to generate some noise, for example by clearing his throat.

\subsection{Security Analysis}
\label{sec:secanalysis}

\noindent\textbf{Remote Attacks.} The security of Sound-Proof stems from the attacker's inability to guess the sound in the victim's environment at the time of the attack.

Let $x$ be the sample recorded by the victim's phone and let $y$ be the sample submitted by the attacker.
A successful impersonation attack requires  the average power of both signals to be above $\tau_{dB}$, and each of the one-third octave band components
of the two signals to be highly correlated.
That is, the two samples must satisfy $Pwr(x)>\tau_{dB}$, $Pwr(y)>\tau_{dB}$ and $S_{x,y}>\tau_C$ with $l<\ell_{max}$.

We bound the lag $l$ between $0$ and $\ell_{max}$ to increase the security of the scheme against an adversary that successfully guesses
the noise in the victim's environment at the time of the attack.
Even if the adversary correctly guesses the noise in the victim's environment and can submit a similar audio sample,
the two samples must be synchronized with an error smaller than $\ell_{max}$.
We also reject audio pairs where either sample has an average power below the threshold $\tau_{dB}$.
This is in order to prevent an impersonation attack when the victim's environment is quiet (e.g., while the victim is sleeping).

Quantifying the entropy of ambient noise, and hence the likelihood of the adversary guessing the signal recorded by the victim's phone, is a challenging task.
Results are dependent on the environment, the language spoken by the victim, his gender or age to cite a few.
In Section~\ref{sec:evaluation} we provide empirical evidence that Sound-Proof can discriminate between legitimate and fraudulent logins, even if the adversary correctly guesses the type of environment where the victim is located.

\noindent\textbf{Co-located Attacks.} Sound-Proof cannot withstand attackers who are co-located with the victim. A co-located attacker can capture the ambient sound in the victim's environment and thus successfully authenticate to the server, assuming that he also knows the victim's password. Sound-Proof shares this limitation with other 2FA mechanisms that do not require the user to interact with his phone and do not assume a secure channel between the phone and the computer (e.g.,~\cite{czeskis12ccs}).
Resistance to co-located attackers requires either a secure phone-to-computer channel (as in~\cite{authy,shirvanian14}) or user-phone interaction (as in~\cite{duosecurity,google2step}).
However, both techniques impose a significant usability burden.

\section{Prototype Implementation}
\label{sec:implementation}

%
%
%
%

Our implementation works with Google Chrome (tested with version 38.0.2125.111), Mozilla Firefox (tested with version 33.0.2) and Opera (tested with version 25.0.1614.68).
We anticipate the prototype to work with different versions of these browsers, as long as they implement the \texttt{navigator.getUserMedia()} API of WebRTC.
We tested the phone application both on Android and on iOS. For Android, on a Samsung Galaxy~S3, a Google Nexus~4 (both running Android version 4.4.4), a Sony Xperia Z3 Compact and a Motorola Nexus 6 (running Android version 5.0.2 and 5.1.1, respectively).
We also tested different iPhone models (iPhone~4, 5 and 6) running iOS version 7.1.2 on the iPhone 4, and iOS version 8.1 on the newer models.
The phone application should work on different phone models and with different OS versions without major modifications.

\noindent\textbf{Web Server and Browser.}
The server component is implemented using the CherryPy~\cite{cherrypy} web framework and MySQL database.
We use WebSocket~\cite{websocketsrfc} to push data from the server to the client.
The client-side (browser) implementation is written entirely in HTML and JavaScript.
Encryption of the audio recording uses AES256 with a fresh symmetric key; the symmetric key is encrypted under the public key of the phone using RSA2048.
We use the HTML5 WebRTC API~\cite{webrtcw3c,webrtc}.
In particular, we use the \texttt{navigator.getUserMedia()} API to access the local microphone from within the browser.
Our prototype does not require browser code modifications or plugins.

\noindent\textbf{Software Token.}
We implement the software token as an Android application as well as an iOS application. The mobile application stays idle in the background and is automatically activated when a push notification arrives. Push messages for Android and iOS use the Google GCM (Google Cloud Messaging) APIs~\cite{gcm} and
Apple's APN (Apple Push Notifications) APIs~\cite{apn} (in particular the silent push notification feature), respectively. Phone to server communication is protected with TLS. 

Most of the Android code is written in Java (Android SDK), while the component that processes the audio samples is written in C (Android NDK).
In particular, we use the ARM Ne10 library, based on the ARM NEON engine~\cite{neon} to optimize vector operations and FFT computations.
The iOS application is written in Objective-C and uses Apple's vDSP package of the Accelerate framework~\cite{accelerate},
in order to leverage the ARM NEON technology for vector operations and FFT computations.
On both mobile platforms we parallelize the computation of the similarity score across the available processor cores.

\noindent\textbf{Time Synchronization.}
Sound-Proof requires the recordings from the phone and the computer to be synchronized.
For this reason, the two devices run a simple time-synchronization protocol (based on the Network Time Protocol~\cite{ntpprotocol}) with the server.
The protocol is implemented over HTTP and allows each device to compute the difference between the local clock and the one of the server.
Each device runs the time-synchronization protocol with the server while it is recording via its microphone.
When recording completes, each device adjusts the timestamp of its sample taking into account the clock difference with the server.

\noindent\textbf{Run-time Overhead.}
We compute the run-time overhead of Sound-Proof when the phone is connected either through WiFi or through the cellular network.
We run 1000 login attempts with a Google Nexus 4 for each connection type, and we measure the time from the moment the user submits his username and password to the time the web server logs the user in. On average it takes 4677ms ($\pm$ 181ms) over WiFi and 4944ms ($\pm$ 233ms) over Cellular to complete the 2FA verification.
Table~\ref{tab:performance} shows the average time and the standard deviation of each operation.
The recording time is set to 3 seconds.
The similarity score is computed over the set of one-third octave bands $B=[50\text{Hz}-4\text{kHz}]$.
(Section~\ref{sec:parameters} discusses the selection of the band set.)
After running the time-synchronization protocol, the resulting clock difference was, on average, 42.47ms ($\pm$ 30.35ms).

\begin{table}[t]
\centering
\scalebox{.9}{
{\tabulinesep=.7mm
\setlength{\tabcolsep}{1.2mm}
\begin{tabu}{|l|r|r|} \hline
\textbf{Operations} & Mean (ms) & Std.Dev. \\ \hline
Recording & 3000 & --- \\ \hline
Similarity score computation & 642 & $171$ \\ \hline
Cryptographic operations & 118 & $15$ \\ \hline
\multicolumn{3}{|l|}{\textbf{Networking}} \\ \hline
WiFi & 978 & $135$ \\ \hline
Cellular & 1243 & $209$ \\ \hline 
\end{tabu}}}
\caption{Overhead of the Sound-Proof prototype. On average it takes 4677ms ($\pm$ 181ms) over WiFi and 4944ms ($\pm$ 233ms) over Cellular to complete the 2FA verification.}
\label{tab:performance}
\end{table}




\section{Evaluation}
\label{sec:evaluation}

\noindent\textbf{Data Collection.}
We used our prototype to collect a large number of audio pairs.
We set up a server that supported Sound-Proof.
Two subjects logged in using Google Chrome\footnote{We used Google Chrome since it is currently the most popular browser~\cite{statcounter}. We have also tested
Sound-Proof with other browsers and have experienced similar performance (see Section~\ref{sec:discussion}).} over 4 weeks.
At each login, the phone and the computer recorded audio through their microphones for 3 seconds.
We stored the two audio samples for post-processing.

Login attempts differed in the following settings.
\emph{Environment:}
an office at our lab with either no ambient noise (labelled as Office) or with the computer playing music (Music);
a living-room with the TV on (TV);
a lecture hall while a faculty member was giving a lecture (Lecture);
a train station (TrainStation);
a cafe (Cafe).
\emph{User activity:} being silent,  talking, coughing, or whistling.
\emph{Phone position:} on a table or a bench next to the user, in the trouser pocket, or in a purse.
\emph{Phone model:} Apple iPhone~5 or Google Nexus~4.
\emph{Computer model:} Mac Book Pro ``Mid 2012'' running OS X10.10 Yosemite or Dell E6510 running Windows 7.

At the end of the 4 weeks we had collected between 5 and 15 login attempts per each setting, totaling
2007 login attempts (4014 audio samples).

\subsection{Analysis}
\label{sec:parameters}
We used the collected samples to find the configuration of system parameters (i.e., $\tau_{dB},\ \ell_{max},\ B,$ and $\tau_C$) that led to
the best results in terms of False Rejection Rate (FRR) and the False Acceptance Rate (FAR).
A false rejection occurs when a legitimate login is rejected.
A false acceptance occurs when a fraudulent login is accepted.
A fraudulent login is accepted if the sample submitted by the attacker
and the sample recorded by the victim's phone have a similarity score greater than $\tau_C$, and if both samples have an average power greater than $\tau_{dB}$.

To compute the FAR, we used the following strategy.
For each phone sample collected by one of the subjects (acting as the victim),
we use all the computer samples collected by the other subject as the attacker's samples. We then switch the roles of the two subjects and repeat the procedure.
The total number of victim--adversary sample pairs we considered was
2,045,680.

\noindent\textbf{System Parameters.}
We set the average power threshold $\tau_{dB}$ to 40dB which, based on our measurements,
is a good threshold to reject silence or very quiet recordings like the sound of a fridge buzzing or the sound of a clock ticking.
Out of 2007 login attempts we found 5 attempts to have an average power of either sample below 40dB and we discard them for the rest of the evaluation.

We set $\ell_{max}$ to 150ms because this was the highest clock difference experienced while testing our time-synchronization protocol (see Section~\ref{sec:implementation}).

%

\begin{figure}[t]
\centering
\includegraphics[width=\columnwidth]{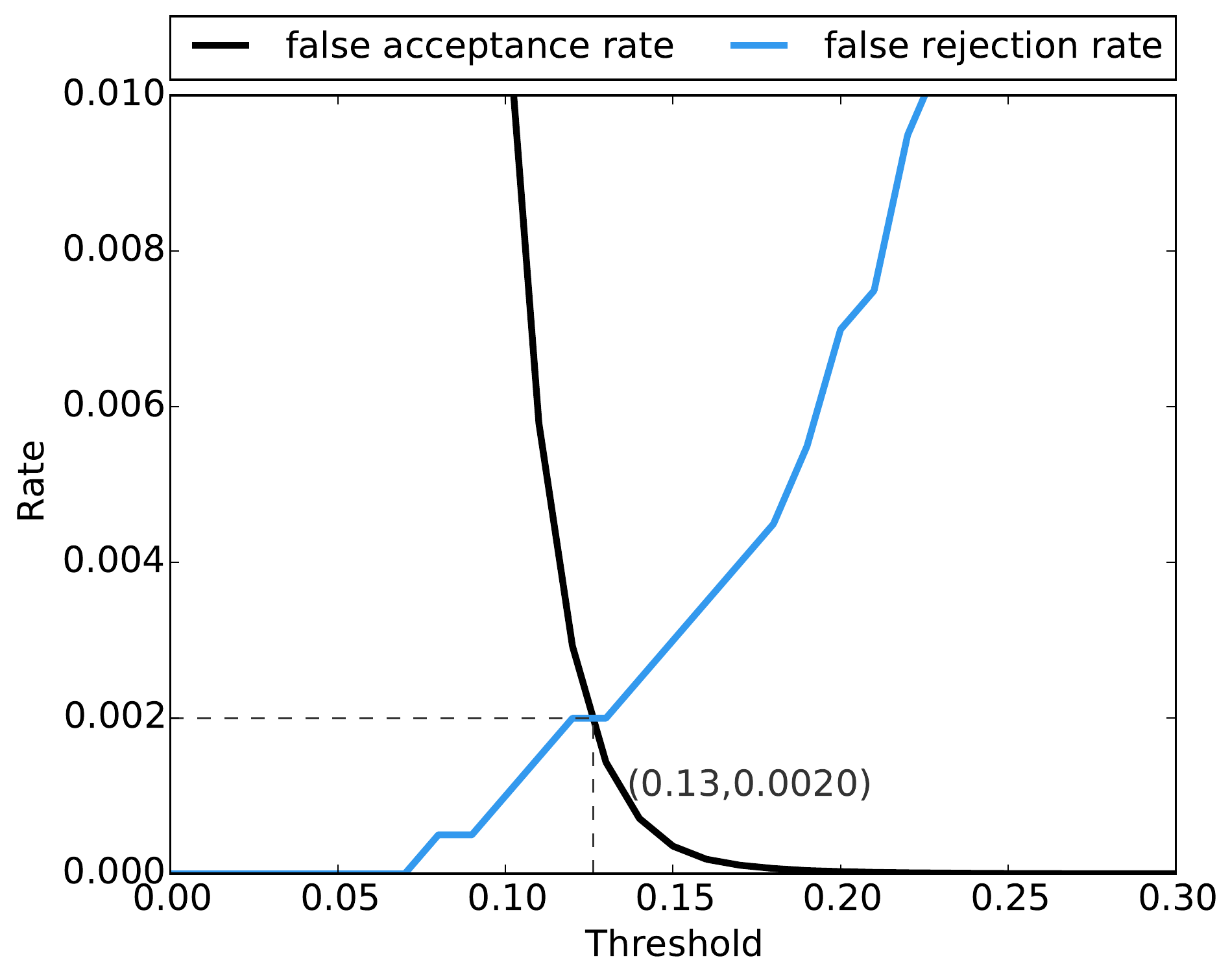}
\caption{False Rejection Rate and False Acceptance Rate as a function of the threshold $\tau_C$ for $B=[50\text{Hz}-4\text{kHz}]$.
The Equal Error Rate is $0.0020$ at $\tau_{C}=0.13.$}
\label{fig:eer_average}
\end{figure}

\begin{figure}[t]
\centering
\subfigure[False Rejection Rate and False Acceptance Rate when usability and security have different weights.] {
\includegraphics[width=\columnwidth]{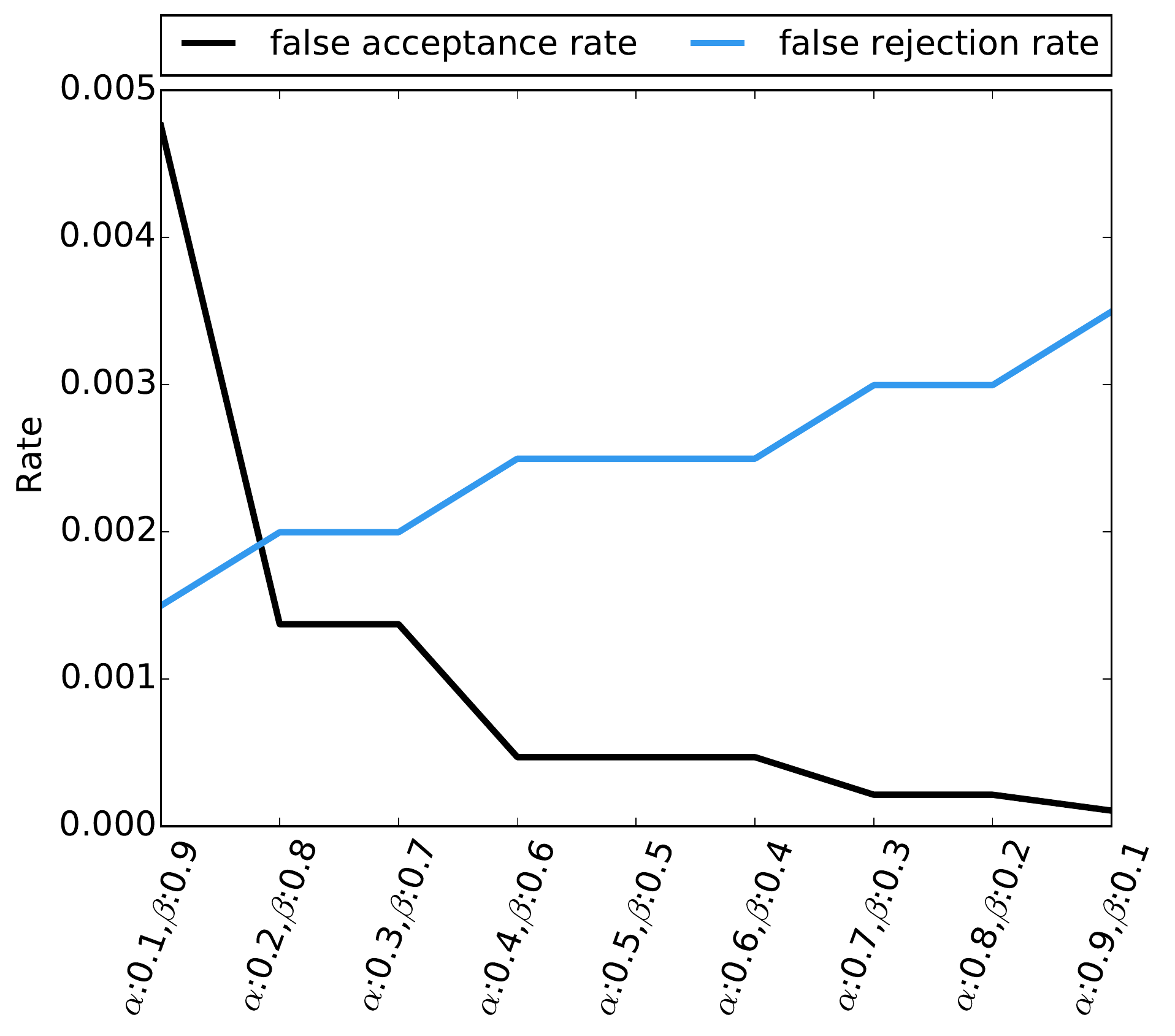}
\label{fig:alfabeta}
}
\subfigure[One-third octave bands and similarity score threshold.] {
\scalebox{.9}{
{\tabulinesep=.7mm
	\setlength{\tabcolsep}{1.2mm}
\begin{tabu}[b]{|l|c|c|}
\hline
			&	B & $\tau_c$ \\
\hline
	$\alpha=0.1,\ \beta=0.9$  & $[80\text{Hz}-2500\text{Hz}]$ & 0.12 \\ \hline
    $\alpha=0.2,\ \beta=0.8$  & $[50\text{Hz}-2500\text{Hz}]$  & 0.14 \\ \hline
    $\alpha=0.3,\ \beta=0.7$  & $[50\text{Hz}-2500\text{Hz}]$ & 0.14 \\ \hline
    $\alpha=0.4,\ \beta=0.6$  & $[50\text{Hz}-800\text{Hz}]$  & 0.19 \\ \hline
    $\alpha=0.5,\ \beta=0.5$  & $[50\text{Hz}-800\text{Hz}]$  & 0.19 \\ \hline
	$\alpha=0.6,\ \beta=0.4$  & $[50\text{Hz}-800\text{Hz}]$  & 0.19 \\ \hline
    $\alpha=0.7,\ \beta=0.3$  & $[50\text{Hz}-1000\text{Hz}]$  & 0.2 \\ \hline
    $\alpha=0.8,\ \beta=0.2$  & $[50\text{Hz}-1000\text{Hz}]$  & 0.2 \\ \hline
    $\alpha=0.9,\ \beta=0.1$  & $[50\text{Hz}-1250\text{Hz}]$ & 0.21 \\ \hline
\end{tabu}}
}

\label{tab:alfabeta}

}

\caption{Minimizing $f=\alpha\cdot FRR+\beta\cdot FAR$, for $\alpha\in[0.1,\ldots,0.9]$ and $\beta=1-\alpha$.}
\end{figure}

\begin{figure}[t]
\centering
\includegraphics[width=\linewidth]{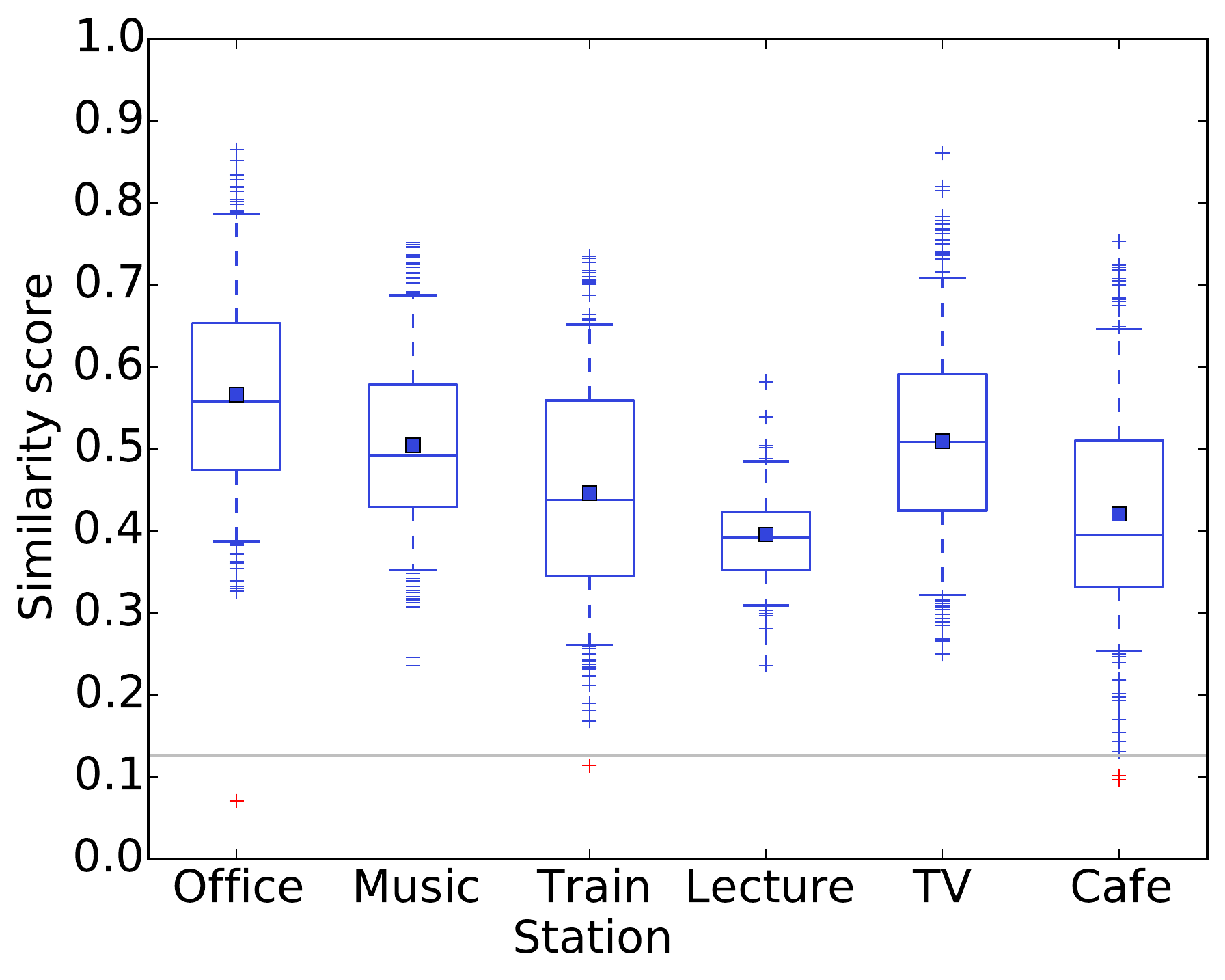}
\caption{Impact of the environment on the False Rejection Rate.}
\label{fig:environment}
\end{figure}

An important parameter of Sound-Proof is the set $B$ of one-third octave bands to consider when computing the similarity score described in Section~\ref{sec:similarity}.
The goal is to select a spectral region that (\emph{i}) includes most common sounds and (\emph{ii}) is robust to attenuation and directionality of audio signals.
We discarded bands below $50\text{Hz}$ to remove very low-frequency noises.
We also discarded bands above $8\text{kHz}$, because these frequencies are attenuated by fabric and they are not suitable for scenarios where the phone is in a pocket or a purse.
We tested all sets of one-third octave bands $B=[x-y]$ where $x$ ranged from $50$Hz to $100$Hz and $y$ ranged from $630$Hz to $8$kHz.

We found the smallest Equal Error Rate (ERR, defined as the crossing point of FRR and FAR) when using $B=[50\text{Hz}-4\text{kHz}]$.
Figure~\ref{fig:eer_average} shows the FRR and FAR using this set of bands where the ERR is $0.0020$ at $\tau_{C}=0.13$.
We experienced worse results with one-third octave bands above 4kHz.
This was likely due to the high directionality of the microphones found on commodity devices when recoding sounds at those frequencies~\cite{book}.

We also computed the best set of one-third octave bands to use in case usability and security are weighted differently by the service provider.\footnote{For example,
a social network provider may value usability higher than security.}
In particular, we computed the sets of bands that minimized $f=\alpha\cdot FRR+\beta\cdot FAR$, for $\alpha\in[0.1,\ldots,0.9]$ and $\beta=1-\alpha$.
Figure~\ref{tab:alfabeta} shows the set of bands that provided the best results for each configuration of $\alpha$ and $\beta$.
As before, we experienced better results with bands below 4kHz.
Figure~\ref{fig:alfabeta} plots the FRR and FAR against the possible values of $\alpha$ and $\beta$.
We stress that the set of bands may differ across two different points on the x-axis.

Experiments in the remaining of this section were run with the configuration of the parameters that minimized the ERR to $0.0020$:
$\tau_{dB}=40\text{dB},\ \ell_{max}=150\text{ms},\ B=[50\text{Hz}-4\text{kHz}],$ and $\tau_C=0.13$.

\subsection{False Rejection Rate}

\begin{figure*}[t]
\centering
%
\subfigure[User Activity]{%
\includegraphics[width=.43\linewidth]{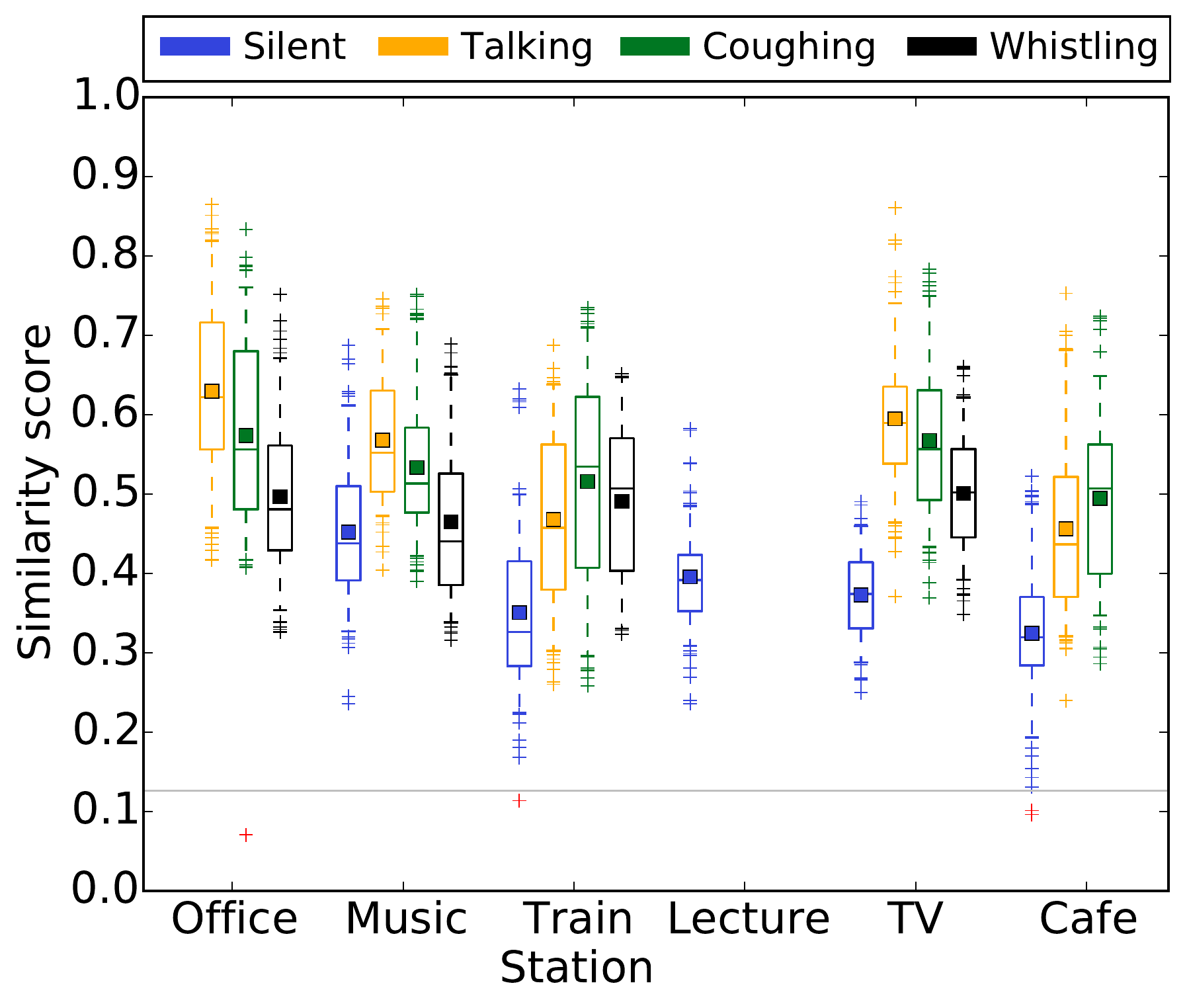}
\label{fig:activity}}
\quad
\subfigure[Phone position]{%
\includegraphics[width=.43\linewidth]{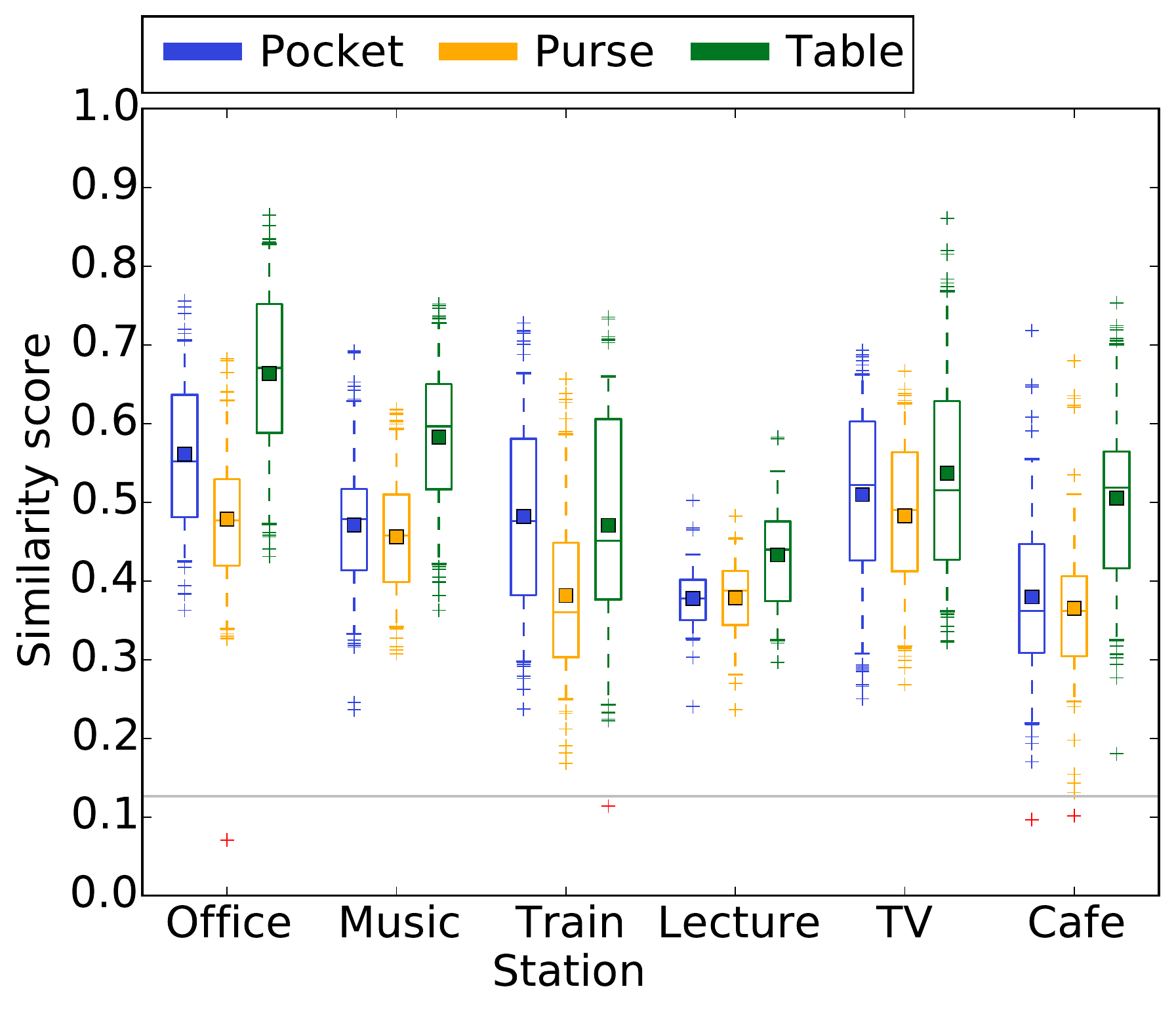}
\label{fig:phoneposition}}
\subfigure[Phone model]{%
\includegraphics[width=.43\linewidth]{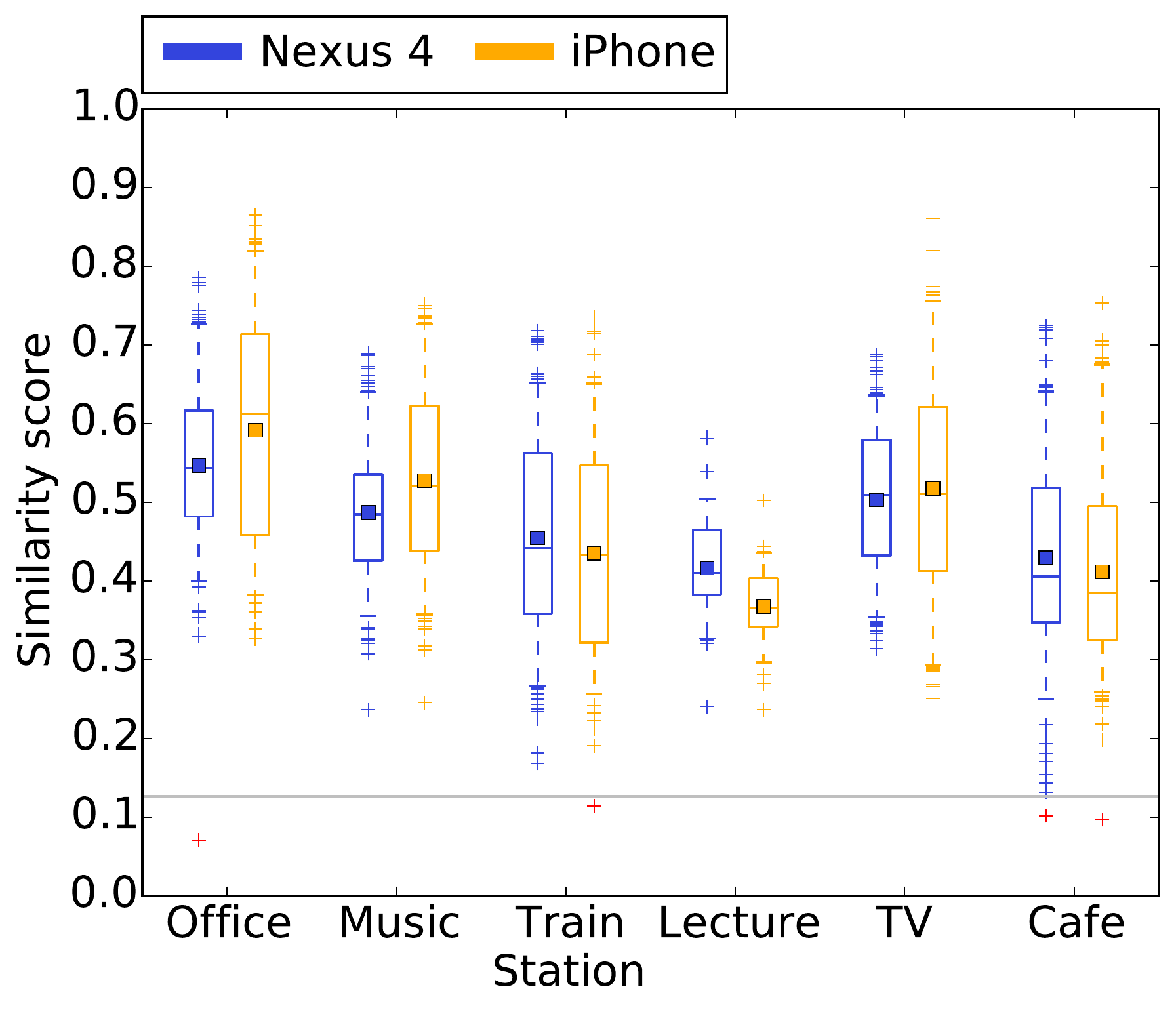}
\label{fig:phone}}
\quad
\subfigure[Computer]{%
\includegraphics[width=.43\linewidth]{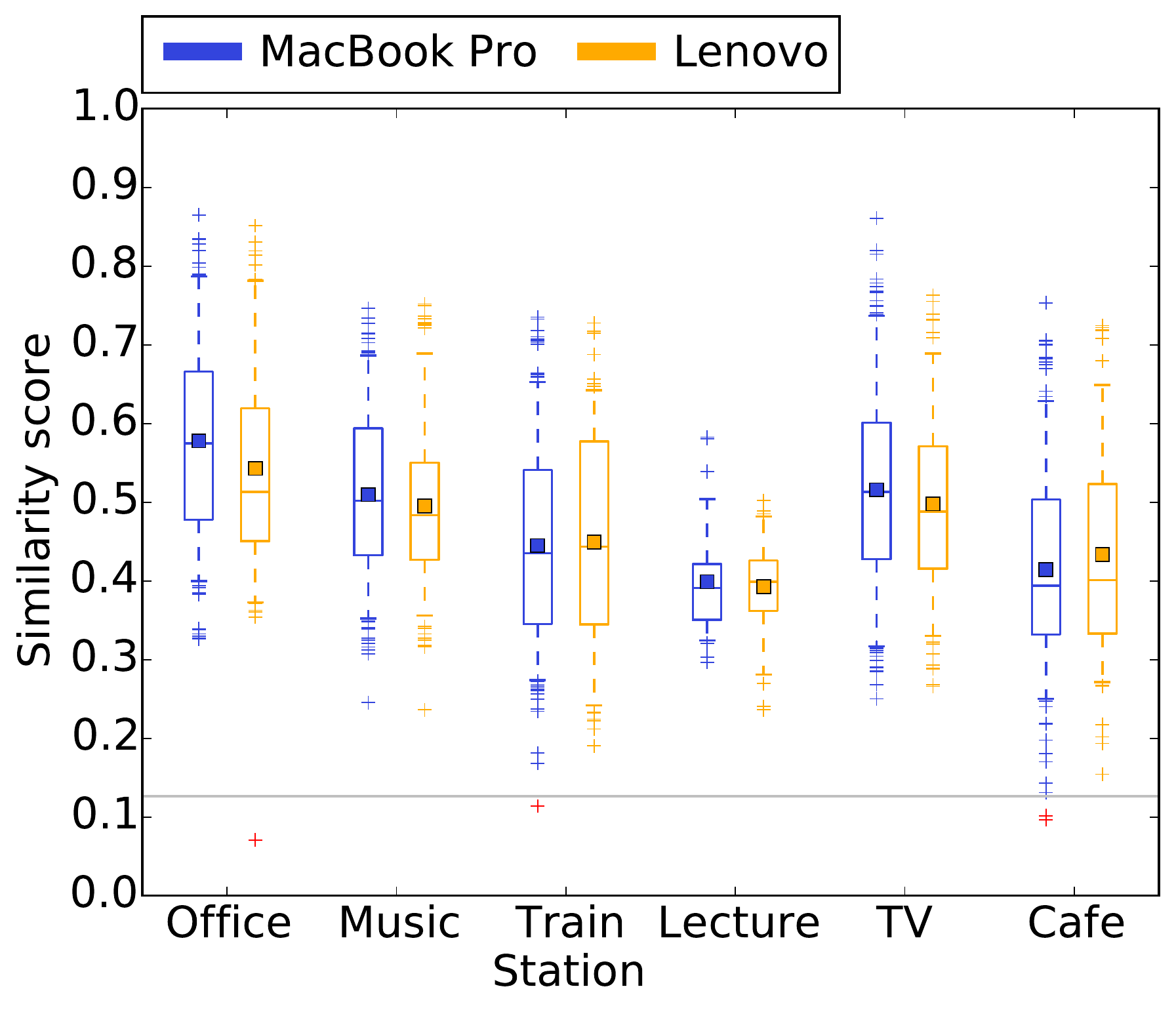}
\label{fig:laptop}}
%
\caption{Impact of user activity, phone position, phone model, and computer model on the False Rejection Rate.}
\label{fig:diffparams}
\end{figure*}

In the following we evaluate the impact of each setting that we consider (environment, user activity, phone position, phone model, and computer model) on the FRR.
Figures~\ref{fig:environment} and~\ref{fig:diffparams} show a box and whisker plot for each setting.
The whiskers mark the 5th and the 95th percentiles of the similarity scores.
The boxes show the 25th and 75th percentiles.
The line and the solid square within each box mark the median and the average, respectively.
A gray line marks the similarity score threshold ($\tau_C=0.13$) and each red dot in the plots denotes a login attempt where the similarity score was below that threshold (i.e., a false rejection).

\noindent\textbf{Environment.}
Figure~\ref{fig:environment} shows the similarity scores for each environment.
Sound-Proof fares equally well indoors and outdoors.
We did not experience rejections of legitimate logins for the Music (over 432 logins), the Lecture (over 122 logins), and the TV (over 430 logins) environments.
The FRR was 0.003 (1 over 310 logins) for Office,
0.003 (1 over 370 logins) for TrainStation, and
0.006 (2 over 338 logins) for Cafe. 

\noindent\textbf{User Activity.}
Figure~\ref{fig:activity} shows the similarity scores for different user activities.
In general, if the user makes any noise the similarity score improves.
We only experienced a few rejections of legitimate logins when the user was silent (TrainStation and Cafe) or when he was coughing (Office).
In the Lecture case the user could only be silent.
We also avoided whistling in the cafe, because this may be awkward for some users.
The FRR was
0.005 (3 over 579 logins) when the user was silent,
0.002 (1 over 529 logins) when the user was coughing,
0 (0 over 541 logins) when the user was speaking, and
0 (0 over 353 logins) when the user was whistling. 

\noindent\textbf{Phone Position.}
Figure~\ref{fig:phoneposition} shows the similarity scores for different phone positions.
Sound-Proof performs slightly better when the phone is on a table or on a bench.
Worse performance when the phone is in a pocket or in a purse are likely due to the attenuation caused by the fabric around the microphone.
The FRR was
0.001 (1 over 667 logins) with the phone on a table,
0.001 (1 over 675 logins) with the phone in a pocket, and
0.003 (2 over 660 logins) with the phone in a purse. 

\noindent\textbf{Phone Model.}
Figure~\ref{fig:phone} shows the similarity scores for the two phones.
The Nexus 4 and the iPhone 5 performed equally good across all environments.
The FRR was
0.002 (2 over 884 logins) with the iPhone~5 and
0.002 (2 over 1118  logins) with the Nexus~4. 

\noindent\textbf{Computer.}
Figure~\ref{fig:laptop} shows the similarity scores for the two computers we used.
We could not find significant differences between their performance.
The FRR was
0.002 (3 over 1299 logins) with the MacBook Pro and
0.001 (1 over 703 logins) with the Dell. 

\noindent\textbf{Distance Between Phone and Computer.}
In some settings (e.g., at home), the user's phone may be away from his computer.
For instance, the user could leave the phone in his bedroom while watching TV or working in another room.
We evaluated this scenario by placing the computer close to the TV in a living-room, and testing Sound-Proof while the phone was away from the computer.
For this set of experiments we used the iPhone 5 and the MacBook Pro.
The average noise level by the TV was measured at 50dB.
We tested 3 different distances: 4, 8 and 12 meters (running 20 login attempts for each distance).
All login attempts were successful (i.e., FRR=0).
We also tried to log in while the phone was in another room behind a closed door, but logins were rejected.

\noindent\textbf{Discussion.}
Based on the above results, we argue that the FRR of Sound-Proof is small enough to be practical for real-world usage. To put it in perspective, the FRR of Sound-Proof is likely to be smaller than the FRR due to mistyped passwords (0.04, as reported in~\cite{kumar07}).

\begin{table}[t]
\centering
\scalebox{.9}{
{\tabulinesep=.7mm
	\setlength{\tabcolsep}{1.2mm}
\begin{tabu}{|l|c|c|c|}
\hline
&\multicolumn{3}{c|}{False Acceptance Rate}\\
\hline
				& SC-SP & SC-DP & DC-DP\\
\hline
	TV channel 1 & 1   & 0.1 & 0.1 \\ \hline 
	TV channel 2 & 1   & 1   & 0 \\ \hline 
	TV channel 3 & 1   & 0   & - \\ \hline 
	TV channel 4 & 1   & 0   & - \\ \hline 
	Web radio 1  & 1   & 0   & 0.4 \\ \hline 
	Web radio 2  & 0.1 & 0.8 & 0.8 \\ \hline 
	Web TV 1     & 0   & 0   & 0 \\ \hline 
	Web TV 2     & 0   & 0   & 0 \\ \hline 
\end{tabu}}
}
\caption{False Acceptance Rate when the adversary and the victim devices record the same broadcast media.
SC-SP stands for ``same city and same Internet/cable provider'',
SC-DP stands for ``same city but different Internet/cable providers'',
DC-DP stands for ``different cities and different Internet/cable providers''.
A dash in the table means that the TV channel was not available at the victim's location.}
\label{tab:tvattack}
\end{table}

\subsection{Advanced Attack Scenarios}
\label{sec:far}
A successful attack requires the adversary to submit a sample that is very similar to the one recoded by the victim's phone.
For example, if the victim is in a cafe, the adversary should submit an audio sample that features typical sounds of that environment.
In the following we assume a strong adversary that correctly guesses the victim's environment.
We also evaluate the attack success rate in scenarios where the victim and the attacker access the same broadcast audio source from different locations.

\noindent\textbf{Similar Environment Attack.}
In this experiment we assume that the victim and the adversary are located in similar environments.
For each environment, we compute the FAR between each phone sample collected by one subject (the victim) and all the computer samples of the other subject (the adversary). We then switch the roles of the two subjects and repeat the procedure.
The FAR  for the Music and the TV environments were 0.012 (1063 over 91960 attempts) and 0.003 (311 over 90992 attempts), respectively.
The FAR for the Lecture environment was  0.001 (8 over 7242 attempts).
When both the victim and the attacker were located at a train station the FAR was 0.001 (44 over 67098 attempts).
The FAR for the Office environment was 0.025 (1194 over 47250 attempts).
When both the victim and the attacker were in a cafe the FAR was 0.001 (32 over 56994 attempts).

The above results show low FAR even when the attacker correctly guesses the victim's environment.
This is due to the fact that ambient noise in a given environment is influenced by random events (e.g., background chatter, music, cups clinking, etc.) that cannot be controlled or predicted by the adversary.

%

\noindent\textbf{Same Media Attack.}
In this experiment we assume that the victim and the adversary access the same audio source from different locations.
This happens, for example, if the victim is watching TV and the adversary correctly guesses the channel to which the victim's TV is tuned.
We place the victim's phone and the adversary's computer in different locations, but each of them next to a smart TV that was also capable of streaming web media.
Since the devices have access to two identical audio sources, the adversary succeeds if the lag between the two audio signals is less than $\ell_{max}$.
We tested 4 cable TV channels, 2 web radios and 2 web TVs.
For each scenario, we run the attack 100 times and report the FAR in  Table~\ref{tab:tvattack}.
When the  victim and the attacker were in the same city, we experienced differences based on the media provider.
When the TVs reproduced content broadcasted by the same provider, the signals were closely synchronized and the similarity score was above the threshold $\tau_C$.
The FAR dropped in the case of web content.
When the TVs reproduced content supplied by different providers, the lag between the signals caused the similarity score to drop below $\tau_C$ in most of the cases.
The similarity score sensibly dropped when the victim and the attacker were located in different cities.


\section{User Study}
\label{sec:userstudy}

The goal of our user study was to evaluate the usability of Sound-Proof and to compare it with the usability of Google 2-Step Verification (2SV), since 2FA based on verification codes is arguably the most popular. (We only considered the version of Google 2SV that uses an application on the user's phone to generate verification codes.)
We stress that the comparison focuses solely on the usability aspect of the two methods.
In particular, we did not make the participants aware of the difference in the security guarantees, i.e., the fact that Google 2SV can better resist co-located attacks.

We ran repeated-measure experiments where each participant was asked to log in to a server using both mechanisms in random order.
After using each 2FA mechanism, participants ranked its usability answering the System Usability Scale (SUS)~\cite{sus}.
The SUS is a widely-used scale to assess the usability of IT systems~\cite{bangor}.
The SUS score ranges from 0 to 100, where higher scores indicate better usability.

\subsection{Procedure}

\noindent\textbf{Recruitment.}
We recruited participants using a snowball sampling method. Most subjects were recruited outside our department and were not working in or studying computer science.
The study was advertised as a user study to ``evaluate the usability of two-factor authentication mechanisms''.
We informed participants that we would not collect any personal information and offered a compensation of CHF 20.
Among all respondents to our email, we discarded the ones that were security experts and ended up with 32 participants.

\noindent\textbf{Experiment.}
The experiment took place in our lab where we provided a laptop and a phone to complete the login procedures. Both devices were connected to the Internet through WiFi.
We set up a Gmail account with Google 2SV enabled.
We also created another website that supported Sound-Proof and mimicked the Gmail UI.

Participants saw a video where we explained the two mechanisms under evaluation.
We told participants that they would need to log in using the account credentials and the hardware we provided.
We also explained that we would record the keystrokes and the mouse movements (this allowed us to time the login attempts).

We then asked participants to fill in a pre-test questionnaire designed to collect demographic information.
Participants logged in to our server using Sound-Proof and to Gmail using Google 2SV.
We randomized the order in which each participant used the two mechanisms.
After each login, participants rated the 2FA mechanism answering the SUS.

At the end of the experiment participants filled in a post-test questionnaire that
covered aspects of the 2FA mechanisms under evaluation not covered by the SUS.

\begin{figure*}[t]
\centering
\subfigure[SUS answers for Sound-Proof]{%
\includegraphics[width=.32\linewidth]{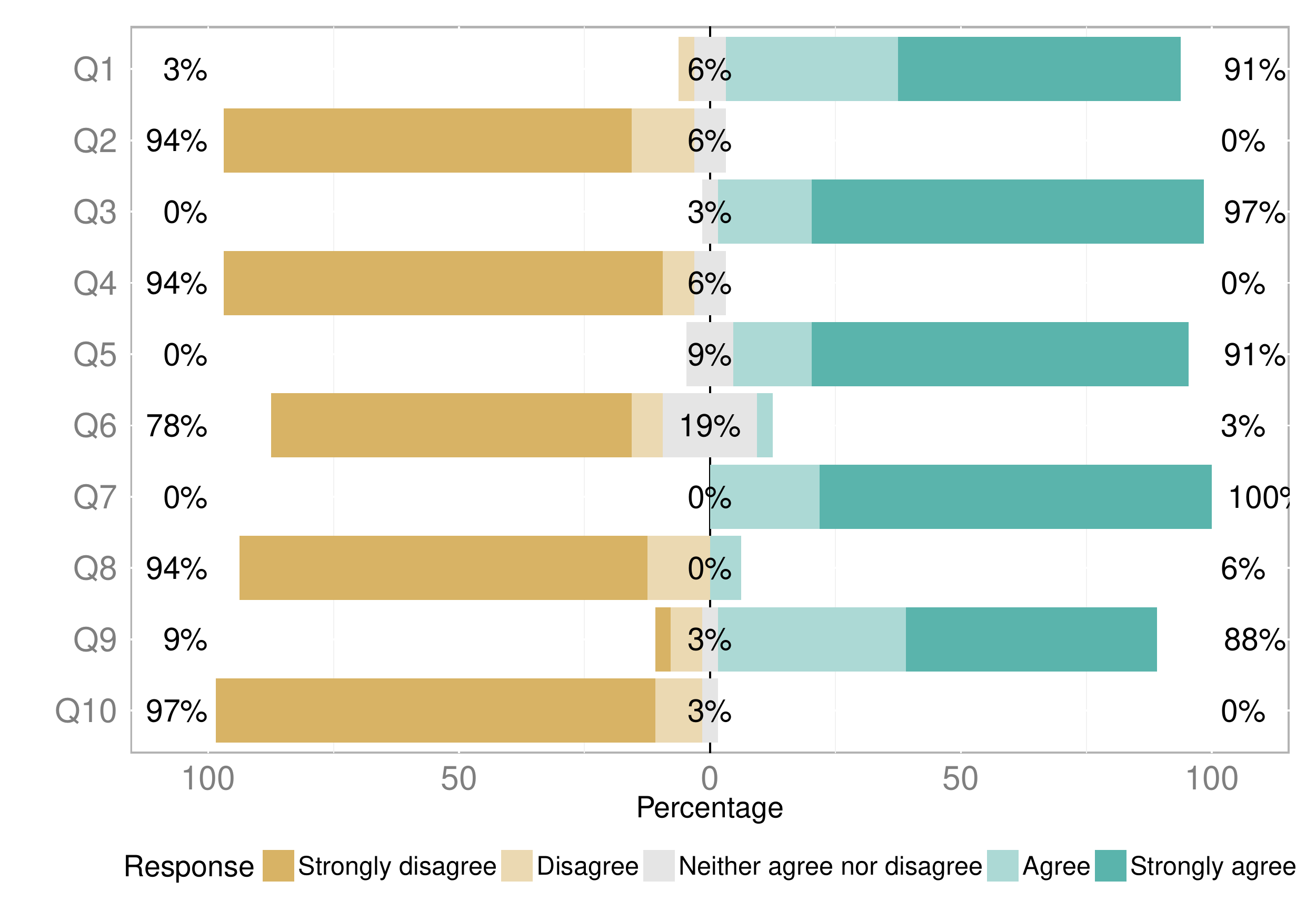}
\label{fig:susaudio}}
\subfigure[SUS answers for Google 2SV]{%
\includegraphics[width=.32\linewidth]{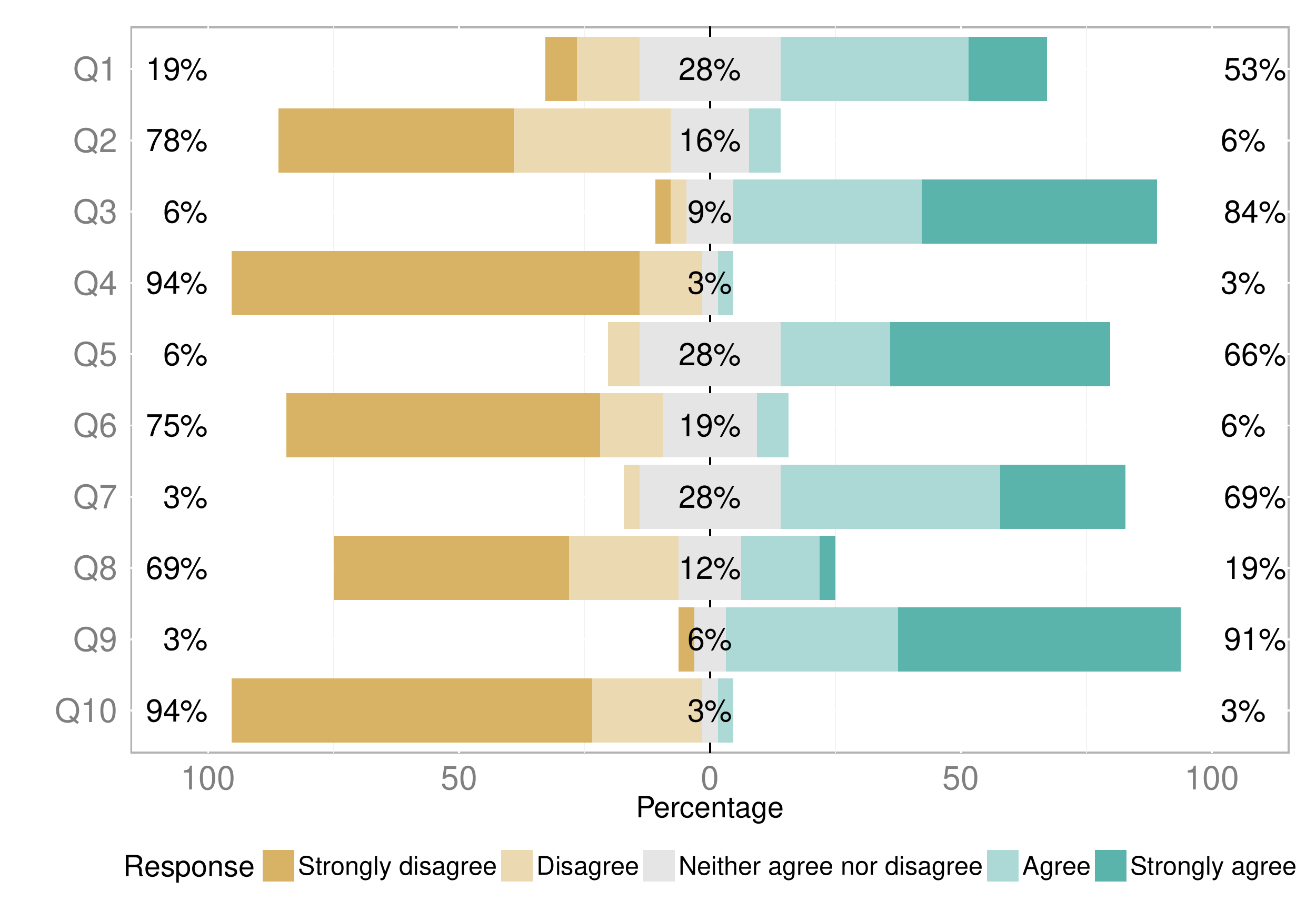}
\label{fig:suscode}}
\subfigure[Answers to the Post-test questionnaire]{%
\includegraphics[width=.32\linewidth]{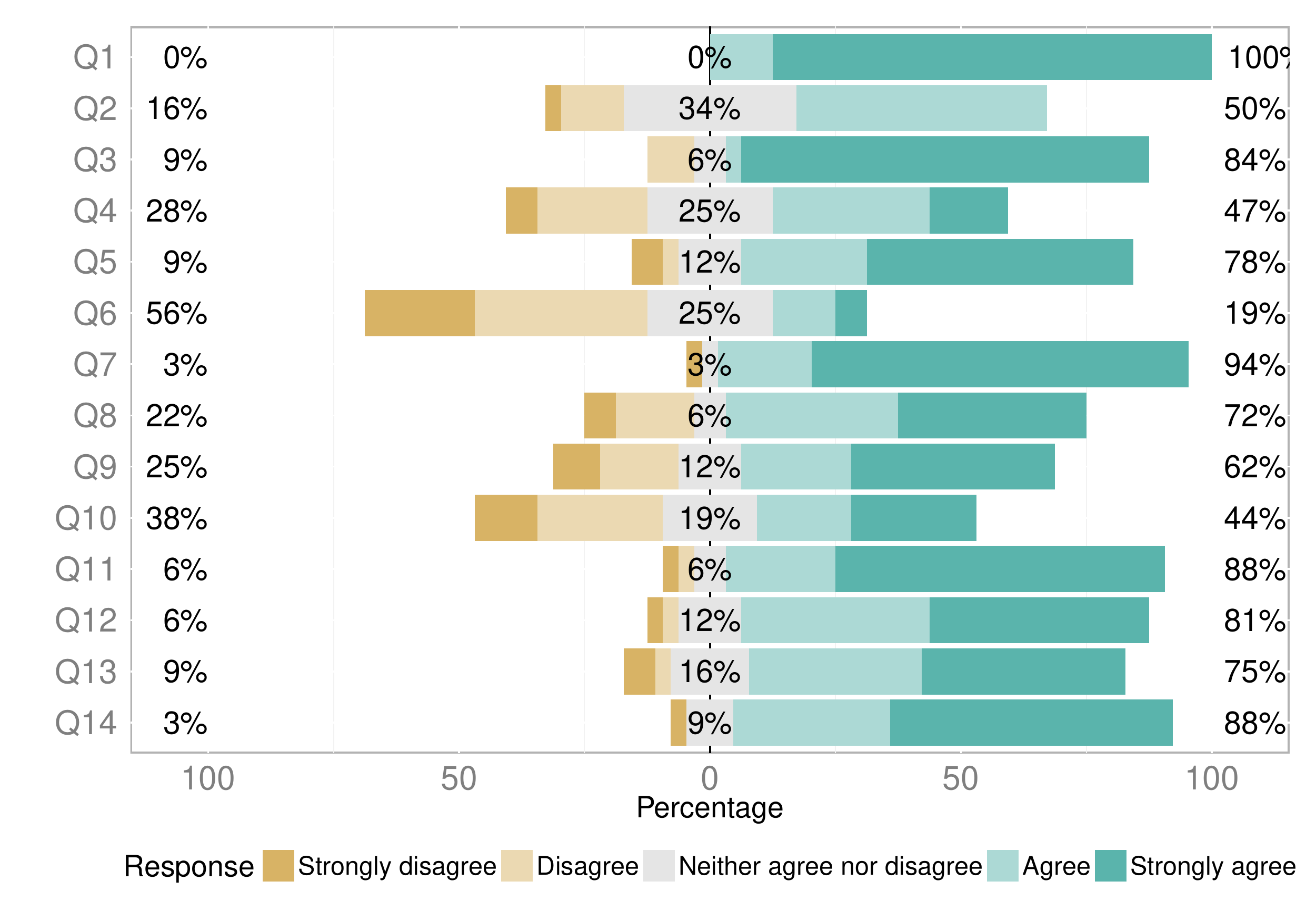}
\label{fig:post}}
\caption{ Distribution of the answers by the participants of the user study.
                System Usability Scale (SUS) of Sound-Proof (a) and Google 2-Step Verification (b), as well as the Post-test questionnaire (c).
            Percentages on the left side include participants that answered ``Strongly disagree'' or ``Disagree''.
            Percentages in the middle account for participants that answered ``Neither agree, nor disagree''.
            Percentages on the right side include participants that answered ``Agree'' or ``Strongly agree''.
}
\label{fig:likerts}
\end{figure*}

\subsection{Results}

\noindent\textbf{Demographics.}
58\% of the participants were between 21 and 30 years old.
25\% of the participants were between 31 and 40 years old.
The remaining 17\% of the participants were above 40 years old.
53\% of the participants were female.
69\% of the participants had a master or doctoral degree.
50\% of the participants used 2FA for online banking and only 13\% used Google 2SV to access their email accounts.

\noindent\textbf{SUS Scores.}
The mean SUS score for Sound-Proof was $91.09$ ($\pm 5.44$).
The mean SUS score for Google 2SV was $79.45$ ($\pm 7.56$).
Figure~\ref{fig:susaudio} and Figure~\ref{fig:suscode} show participant answers on 5-point Likert-scales for Sound-Proof and for Google 2SV, respectively. To analyze the statistical significance of these results, we used the following null hypothesis:
``there will be no difference in perceived usability between Sound-Proof and Google 2SV''.
A one-way ANOVA test revealed that the difference of the SUS scores was statistically significant
($F(1,31) = 21.698,\ p<.001,\ \eta_p^2 = .412$), thus the null hypothesis can be rejected.
We concluded that users perceive Sound-Proof to be more usable than Google 2SV.
Appendix~\ref{app:sus} reports the items of the SUS.

\noindent\textbf{Login Time.}
We measured the login time from the moment when a participant clicked on the ``login'' button (right after entering the password),
to the moment when that participant was logged in.
We neglected the time spent entering username and password because we wanted to focus only on the time required by the 2FA mechanism.
Login time for Sound-Proof was $4.7$ seconds ($\pm 0.2$ seconds); this time was required for the phone to receive the computer's sample and compare it with the one recorded locally. 
With Google 2SV, login time increased to $24.4$ seconds ($\pm 7.1$ seconds); this time was required for the participant to take the phone, start the
application and copy the verification code from the phone to the browser.

\noindent\textbf{Failure Rates.}
We did not witness any login failure for either of the two methods.
We speculate that this may be due to the priming of the users right before the experiment,
when we explained how the two methods work and that Sound-Proof may require users to make some noise in quiet environments.

\noindent\textbf{Post-test Questionnaire.}
The post-test questionnaire was designed to collect information on the perceived quickness of the two mechanisms (Q1--Q2)
and participants willingness to adopt any of the schemes (Q3--Q6).
We also included items to inquire if participants would feel comfortable using the mechanisms in different environments (Q7--Q14).
Figure~\ref{fig:post} shows participants answers on 5-point Likert-scales.
The full text of the items can be found in Appendix~\ref{app:posttest}.

All participants found Sound-Proof to be quick (Q1), while only 50\% of the participants found Google 2SV to be quick (Q2).
If 2FA were mandatory, 84\% of the participants would use Sound-Proof (Q3) and 47\% would use Google 2SV (Q4).
In case 2FA were optional the percentage of participants willing to use the two mechanisms dropped to 78\% for Sound-Proof (Q5) and to 19\% for Google 2SV (Q6).
Similar to~\cite{petsas15eurosec,imperi13umsurvey}, our results for Google 2SV suggest that users are likely not to use 2FA if it is optional.
With Sound-Proof, the difference in user acceptance between a mandatory and an optional scenario is only 6\%.

We asked participants if they would feel comfortable using either mechanism at home, at their workplace, in a cafe, and in a library.
95\% of the participants would feel comfortable using Sound-Proof at home (Q7) and 77\% of the participants would use it at the workplace (Q8).
68\% would use it in a cafe (Q9) and 50\% would use it in a library (Q10).
Most participants (between 91\% and 82\%) would feel comfortable using Google 2SV in any of the scenario we considered (Q11--Q14).

The results of the post-test questionnaire suggest that users may be willing to adopt Sound-Proof because it is quicker and causes less burden, compared to Google 2SV.
In some public places, however, users may feel more comfortable using Google 2SV. In Section~\ref{sec:discussion} we discuss how to integrate the two approaches.

The post-test questionnaire allowed participants to comment on the 2FA mechanisms evaluated.
Most participants found Sound-Proof to be user-friendly and appreciated the lack of interaction with the phone.
Appendix~\ref{app:comments} lists some of the users' comments.


\section{Discussion}
\label{sec:discussion}

\noindent\textbf{Software and Hardware Requirements.}
Similar to any other 2FA based on software tokens, Sound-Proof requires an application on the user's phone.
Sound-Proof, however, does not require additional software on the computer and seamlessly works with any
HTML5-compliant browser that implements the WebRTC API.
Chrome, Firefox and Opera, already support WebRTC and a version of Internet Explorer supporting WebRTC will soon be released~\cite{iewebrtc}.
Sound-Proof needs the phone to have a data connection. Moreover, both the phone and the computer where the browser is running must be equipped with a microphone. Microphones are ubiquitous in phones, tablets and laptops. If a computer such as a desktop machine does not have an embedded microphone, Sound-Proof requires an external microphone, like the one of a webcam.

\noindent\textbf{Other Browsers.}
Section~\ref{sec:evaluation} evaluates Sound-Proof using Google Chrome.
We have also tested Sound-Proof with Mozilla Firefox and Opera.
Each browser may use different algorithms to process the recorded audio (e.g., filtering for noise reduction), before delivering it to the web application.
The WebRTC specification does not yet define how the recorded audio should be processed, leaving the specifics of the implementation to the browser vendor. When we ran our tests, Opera behaved like Chrome.
Firefox audio processing was slightly different and it affected the performance our prototype.
In particular, the Equal Error Rate computed over the samples collected while using Firefox was 0.012.
We speculate that a better Equal Error Rate can be achieved with any browser if the software token performs the same audio processing of the browser being used to log in.

\noindent\textbf{Privacy.}
The noise in the user's environment may leak private information to a prying server. In our design, the audio recorded by the phone is never uploaded to the server.
A malicious server can also access the computer's microphone while the user is visiting the server's webpage.
This is already the case for a number of websites that require access to the microphone.
For example, websites for language learning, Gmail (for video-chats or phone calls), live chat-support services, or any site
that uses speech-recognition require access to the microphone and may record the ambient noise any time the user visits the provider's webpage.
All browsers we tested ask the user for permission before allowing a website to use \texttt{getUserMedia}. Moreover, browsers show an alert when a website triggers recording from the microphone. Providers are likely not to abuse the recording capability, since their reputation would be affected, if users detect unsolicited recording.

\noindent\textbf{Quiet Environments.}
Sound-Proof rejects a login attempt if the power of either sample is below $\tau_{dB}$.
In case the environment is too quiet, the website can prompt the user to make any noise (by, e.g., clearing his throat, knocking on the table, etc.).


\noindent\textbf{Fallback to Code-based 2FA.}
Sound-Proof can be combined with 2FA mechanisms based on verification codes, like Google 2SV.
For example, the webpage can employ Sound-Proof as the default 2FA mechanism, but give to the user the option to log in entering a verification code.
This may be useful in cases where the environment is quiet and the user feels uncomfortable making noise.
Login based on verification codes is also convenient when the phone has no data connectivity (e.g., when roaming).

\noindent\textbf{Failed Login Attempts and Throttling.}
Sound-Proof deems a login attempt as fraudulent if the similarity score between the two samples is below the threshold $\tau_C$ or if the power of either sample is below $\tau_{dB}$.
In this case, the server may request the two devices to repeat the recording and comparison phase.
After a pre-defined number of failed trials, the server can fall-back to a 2FA mechanism based on verification codes.
The server can also throttle login attempts in order to prevent ``brute-force'' attacks and to protect the user's phone battery from draining.

\noindent\textbf{Login Evidence.}
Since audio recording and comparison is transparent to the user, he has no means to detect an ongoing attack. To mitigate this, at each login attempt the phone may vibrate, light up, or display a message to notify the user that a login attempt is taking place.
The Sound-Proof application may also keep a log of the login attempts.
Such techniques can help to make the user aware of fraudulent login attempts.
Nevertheless, we stress that the user does not have to attend to the phone during legitimate login attempts.

\noindent\textbf{Continuous Authentication.}
Sound-Proof can also be used as a form of continuous authentication. The server can periodically trigger Sound-Proof, while the user is logged in and interacts with the website. If the recordings of the two devices do not match, the server can forcibly log the user out. Nevertheless, such use can have a more significant impact on the user's privacy, as well as affect the battery life of the user's phone.

\noindent\textbf{Alternative Devices.}
Our 2FA mechanism uses the phone as a software token.
Another option is to use a smartwatch and we plan to develop a Sound-Proof application for smartwatches based on Android Wear and Apple Watch.
We speculate that smartwatches can further lower the false rejection rate because of the proximity of the computer and the smartwatch during logins.

\noindent\textbf{Logins from the Phone.}
If a user tries to log in from the same device where the Sound-Proof application is running,
the browser and the application will capture audio through the same microphone and, therefore, the login attempt will be accepted. This requires the mobile OS to allow access to the microphone by the browser and, at the same time, by the Sound-Proof application.
If the mobile OS does not allow concurrent access to the microphone, Sound-Proof can fall back to code-based 2FA.

\begin{table*}[t]
\centering
\scalebox{0.8}{
    \begin{tabular}{l|cccccccc|cccccc|ccccccccccc}
        &\multicolumn{8}{c}{Usability}& \multicolumn{6}{|c}{Deployability}&\multicolumn{11}{|c}{Security}\\
        Scheme & \rot{Memorywise-Effortless} & \rot{Scalable-for-Users}&\rot{Nothing-to-Carry} & \rot{Physically Effortless} & \rot{Easy-to-Learn} &\rot{Efficient-to-Use} &\rot{Infrequent-Errors}&\rot{Easy-Recovery-from-Loss}&\rot{Accessible} & \rot{Negligible-Cost-per-User} &\rot{Server-Compatible}&\rot{Browser-Compatible}&\rot{Mature}&\rot{Non-Proprietary} & \rot{Resilient-to-Physical-Observation}&\rot{Resilient-to-Targeted-Impersonation}&\rot{Resilient-to-Throttled-Guessing}&\rot{Resilient-to-Unthrottled-Guessing} &        \rot{Resilient-to-Internal-Observation}&\rot{Resilient-to-Leaks-from-Other-Verifiers}&\rot{Resilient-to-Phishing}&\rot{Resilient-to-Theft}&\rot{No-Trusted-Third-Party}&        \rot{Requiring-Explicit-Consent} &\rot{Unlinkable}\\
 \hline
  Sound-Proof &   &   & S &   & Y & Y & S & S    & Y & Y &   & Y &   & Y    & S &   & Y & Y &  & Y & Y & Y & Y & Y &\\
 \hline
  Google 2SV  &   &   & S &   & Y & S & S & S    & S & S &   & Y & Y &  Y    & S & S & Y & Y &   & Y & Y & Y & Y & Y &\\
 \hline
 PhoneAuth   &   &   & S &   & Y & Y & S & S    & Y & Y &   &   &   & Y    & S &   & Y & Y &   & Y & Y & Y & Y & Y &\\
 \hline
 FBD-WF-WF   &   &   & S &   & Y & S & S & S    & Y & Y &   &   &   & Y    & S &  S & Y & Y &   & Y & Y & Y & Y & Y &\\
 \hline
    \end{tabular}
    }
    \caption{Comparison of Sound-Proof against Google 2-Step Verification (Google 2SV), PhoneAuth~\cite{czeskis12ccs}, and FBD-WF-WF~\cite{shirvanian14}, using the framework of Bonneau et al.~\cite{bonneau12sp}. We use `Y' to denote that the benefit is provided and `S' to denote that the benefit is somewhat provided.\label{tab:comparison}}
\end{table*}

\noindent\textbf{Comparative Analysis.}
We use the framework of Bonneau et al.~\cite{bonneau12sp} to compare Sound-Proof with Google 2-Step Verification (Google 2SV), with PhoneAuth~\cite{czeskis12ccs}, and with  the 2FA protocol of~\cite{shirvanian14} that uses WiFi to create a channel between the phone and the computer (referred to as FBD-WF-WF in~\cite{shirvanian14}).
The framework of  Bonneau et al. considers 25 ``benefits'' that an authentication scheme should provide, categorized in terms of usability, deployability, and security.
Table~\ref{tab:comparison} shows the overall comparison.
The evaluation of Google 2SV in Table~\ref{tab:comparison} matches the one reported by~\cite{bonneau12sp}, besides the fact that we consider Google 2SV to be non-proprietary.
\emph{Usability:}
No scheme is scalable nor it is effortless for the user because they all require a password as the first authentication factor.
They are all ``Quasi-Nothing-to-Carry'' because they leverage the user's phone.
Sound-Proof and PhoneAuth are more efficient to use than Google 2SV because they do not require the user to interact with his phone. They are also more efficient to use than  FBD-WF-WF, because the latter requires a non-negligible setup time every time the user logs in from a new computer.
All mechanisms incur some errors if the user enters the wrong password (Infrequent-Errors).
All mechanisms also require similar recovery procedures if the user loses his phone.
\emph{Deployability:}
Sound-Proof, PhoneAuth, and FBD-WF-WF score better than Google 2SV in the category ``Accessible'' because the user is asked nothing but his password.
The three schemes are also better than Google 2SV in terms of cost per user, assuming users already have a phone.
None of the mechanisms is server-compatible. Sound-Proof and Google 2SV are the only  browser-compatible mechanisms as they require no changes to current browsers or computers.
Google 2SV is more mature, and all of them are non-proprietary.
\emph{Security:}
The security provided by Sound-Proof, PhoneAuth, and FBD-WF-WF is similar to the one provided by Google 2SV.
However, we rate Sound-Proof and PhoneAuth as not resilient to targeted impersonation, since a targeted, co-located attacker can launch the attack from the victim's environment. FBD-WF-WF uses a paired connection between the user's computer and phone, and can better resist such attacks.

\section{Related Work}
\label{sec:relatedwork}

Section~\ref{sec:alternative} discusses  alternative approaches to 2FA. In the following we review related work that leverages audio to verify the proximity of two devices.
Halevi et al.,~\cite{halevi12esorics} use ambient audio to detect the proximity of two devices to thwart relay attacks in NFC payment systems.
They compute the cross-correlation between the audio recorded by the two devices and employ machine-learning techniques to tell whether
the two samples were recorded at the same location or not.
The authors claim perfect results (0 false acceptance and false rejection rate).
They, however, assume the two devices to have the same hardware (the experiment campaign used two Nokia N97 phones).
Furthermore, their setup allows a maximum distance of 30 centimeters between the two devices.
Our application scenario (web authentication) requires a solution that works (\emph{i}) with heterogeneous devices, (\emph{ii}) indoors and outdoors, and (\emph{iii}) irrespective of the
phone's position (e.g., in the user's pocket or purse).
As such, we propose a different function to compute the similarity of the two samples, which we empirically found to be more robust, than what proposed in~\cite{halevi12esorics}, in our settings.

Truong et al.,~\cite{truong14percom} investigate relay attacks in zero-interaction authentication systems and use techniques similar to the ones of~\cite{halevi12esorics}.
They propose a framework that detects co-location of two devices comparing features from multiple sensors, including GPS, Bluetooth, WiFi and audio.
The authors conclude that an audio-only solution is not robust to detect co-location (20\% of false rejections) and advocate for the combination of multiple sensors.
Furthermore, their technique requires the two devices to sense the environment for 10 seconds. This time budget may not be available for web authentication.

The authors of~\cite{schurmannS13tmc} use ambient audio to derive a pair-wise cryptographic key between two co-located devices.
They use an audio fingerprinting scheme similar to the one of~\cite{haitsma02} and leverage fuzzy commitment schemes to accommodate for the difference of the two
recordings. Their scheme may, in principle, be used to verify proximity of two devices in a 2FA mechanism. 
However, the experiments of~\cite{schurmannS13tmc} reveal that
the key derivation is hardly feasible in outdoor scenarios. Our scheme takes advantage of noisy environments and, therefore, can be used in outdoor scenarios like train stations.


\section{Conclusion}
\label{sec:conclusion}
We proposed Sound-Proof, a two-factor authentication mechanism that does not require the user to interact with his phone and that can already be used with major browsers.
We have shown that Sound-Proof works even if the phone is in the user's pocket or purse, and that it fares well both indoors and outdoors.
Participants of a user study rated Sound-Proof to be more usable than Google 2-Step Verification. More importantly, most participants would use Sound-Proof for online services in which 2FA is optional. Sound-Proof improves the usability and deployability of 2FA and, as such, can foster large-scale adoption.


\section*{Acknowledgments}

We thank Kurt Heutschi for the valuable discussions and insights on audio processing. We also thank our shepherd Joseph Bonneau, as well as the anonymous reviewers who helped to improve this paper with their useful feedback and comments.

{\footnotesize
 \bibliographystyle{acm}
\bibliography{paper}

\begin{thebibliography}{10}

\bibitem{accelerate}
{\sc {Apple}}.
\newblock Accelerate framework reference.
\newblock \url{https://goo.gl/WtnCOk}.

\bibitem{apn}
{\sc {Apple}}.
\newblock {Apple Push Notification Service}.
\newblock \url{https://goo.gl/t8UUMf}.

\bibitem{arentz11ubicomp}
{\sc Arentz, W.~A., and Bandara, U.}
\newblock Near ultrasonic directional data transfer for modern smartphones.
\newblock In {\em 13th International Conference on Pervasive and Ubiquitous
  Computing\/} (2011), UbiComp '11.

\bibitem{neon}
{\sc {ARM}}.
\newblock {ARM NEON}.
\newblock \url{http://www.arm.com/products/processors/technologies/neon.php}.

\bibitem{authy}
{\sc {Authy Inc.}}
\newblock {Authy}.
\newblock \url{https://www.authy.com}.

\bibitem{backes09sp}
{\sc Backes, M., Chen, T., D{\"{u}}rmuth, M., Lensch, H. P.~A., and Welk, M.}
\newblock Tempest in a teapot: Compromising reflections revisited.
\newblock In {\em IEEE Symposium on Security and Privacy\/} (2009), SP '09.

\bibitem{backes08sp}
{\sc Backes, M., D{\"{u}}rmuth, M., and Unruh, D.}
\newblock Compromising reflections-or-how to read {LCD} monitors around the
  corner.
\newblock In {\em IEEE Symposium on Security and Privacy\/} (2008), SP '08.

\bibitem{baluja08pr}
{\sc Baluja, S., and Covell, M.}
\newblock Waveprint: Efficient wavelet-based audio fingerprinting.
\newblock {\em Pattern Recognition 41}, 11 (2008), 3467--3480.

\bibitem{bangor}
{\sc Bangor, A., Kortum, P.~T., and Miller, J.~T.}
\newblock An empirical evaluation of the system usability scale.
\newblock {\em International Journal of Human-Computer Interaction 24}, 6
  (2008).

\bibitem{bonneau12sp}
{\sc Bonneau, J., Herley, C., van Oorschot, P.~C., and Stajano, F.}
\newblock The quest to replace passwords: {A} framework for comparative
  evaluation of web authentication schemes.
\newblock In {\em {IEEE} Symposium on Security and Privacy\/} (2012), SP '12.

\bibitem{sus}
{\sc Brooke, J.}
\newblock {SUS - {A} quick and dirty usability scale}.
\newblock {\em Usability evaluation in industry 189}, 194 (1996), 4--7.

\bibitem{imperi13umsurvey}
{\sc {BusinessWire}}.
\newblock Impermium study unearths consumer attitudes toward internet security.
\newblock \url{http://goo.gl/NsUCL7}, 2013.

\bibitem{chandrasekhar11ismir}
{\sc Chandrasekhar, V., Sharifi, M., and Ross, D.~A.}
\newblock Survey and evaluation of audio fingerprinting schemes for mobile
  query-by-example applications.
\newblock In {\em 12th International Society for Music Information Retrieval
  Conference\/} (2011), ISMIR '11.

\bibitem{czeskis12ccs}
{\sc Czeskis, A., Dietz, M., Kohno, T., Wallach, D.~S., and Balfanz, D.}
\newblock Strengthening user authentication through opportunistic cryptographic
  identity assertions.
\newblock In {\em {ACM Conference on Computer and Communications Security}\/}
  (2012), CCS '12.

\bibitem{webrtcw3c}
{\sc {Daniel C. Burnett and Adam Bergkvist and Cullen Jennings and Anant
  Narayanan}}.
\newblock {Media Capture and Streams (W3C Working Draft)}.
\newblock \url{http://www.w3.org/TR/mediacapture-streams/}.

\bibitem{duosecurity}
{\sc {Duo Security, Inc.}}
\newblock {Duo Push}.
\newblock \url{https://www.duosecurity.com/product/methods/duo-mobile}.

\bibitem{rsa}
{\sc {EMC Inc.}}
\newblock {RSA SecurID}.
\newblock \url{https://www.emc.com/security/rsa-securid.htm/}.

\bibitem{encap}
{\sc {Encap Security}}.
\newblock {Encap Security}.
\newblock \url{https://www.encapsecurity.com/}.

\bibitem{websocketsrfc}
{\sc Fette, I., and Melnikov, A.}
\newblock {The WebSocket protocol (RFC 6455)}.
\newblock \url{http://tools.ietf.org/html/rfc6455}, 2011.

\bibitem{fido}
{\sc {FIDO Alliance}}.
\newblock {Fido U2F specifications}.
\newblock \url{https://fidoalliance.org/specifications/}.

\bibitem{gcm}
{\sc {Google}}.
\newblock {Google Cloud Messaging for Android}.
\newblock \url{https://developer.android.com/google/gcm/index.html}.

\bibitem{google2step}
{\sc {Google Inc.}}
\newblock {Google 2-Step Verification}.
\newblock \url{https://www.google.com/landing/2step/}.

\bibitem{slicklogin}
{\sc {Google Inc.}}
\newblock {SlickLogin}.
\newblock \url{http://www.slicklogin.com/}.

\bibitem{webrtc}
{\sc {Google Inc.}}
\newblock {WebRTC}.
\newblock \url{http://www.webrtc.org/}.

\bibitem{gunson11cs}
{\sc Gunson, N., Marshall, D., Morton, H., and Jack, M.~A.}
\newblock User perceptions of security and usability of single-factor and
  two-factor authentication in automated telephone banking.
\newblock {\em Computers \& Security 30}, 4 (2011), 208--220.

\bibitem{haitsma02}
{\sc Haitsma, J., Kalker, T., and Oostveen, J.}
\newblock An efficient database search strategy for audio fingerprinting.
\newblock In {\em 5th Workshop on Multimedia Signal Processing\/} (2002), MMSP
  '02.

\bibitem{halevi12esorics}
{\sc Halevi, T., Ma, D., Saxena, N., and Xiang, T.}
\newblock Secure proximity detection for {NFC} devices based on ambient sensor
  data.
\newblock In {\em 17th European Symposium on Research in Computer Security\/}
  (2012), ESORICS '12.

\bibitem{hazas02ubicomp}
{\sc Hazas, M., and Ward, A.}
\newblock A novel broadband ultrasonic location system.
\newblock In {\em 4th International Conference on Pervasive and Ubiquitous
  Computing\/} (2002), UbiComp.

\bibitem{karapanos14usenix}
{\sc Karapanos, N., and Capkun, S.}
\newblock On the effective prevention of {TLS} man-in-the-middle attacks in web
  applications.
\newblock In {\em 23rd {USENIX} Security Symposium\/} (2014), USENIX Sec '14.

\bibitem{kumar07}
{\sc Kumar, M., Garfinkel, T., Boneh, D., and Winograd, T.}
\newblock Reducing shoulder-surfing by using gaze-based password entry.
\newblock In {\em 3rd Symposium on Usable Privacy and Security\/} (2007), SOUPS
  '07.

\bibitem{iewebrtc}
{\sc {Microsoft}}.
\newblock Bringing interoperable real-time communications to the web.
\newblock
  \url{http://blogs.skype.com/2014/10/27/bringing-interoperable-real-time-communications-to-the-web/}.

\bibitem{firefoxgeolocation}
{\sc {Mozzilla}}.
\newblock {Location-Aware Browsing}.
\newblock \url{https://www.mozilla.org/en-US/firefox/geolocation/}.

\bibitem{ntpprotocol}
{\sc {Network Time Foundation}}.
\newblock {NTP: The Network Time Protocol}.
\newblock \url{http://www.ntp.org/}.

\bibitem{owasp_mitb}
{\sc {OWASP}}.
\newblock Man-in-the-browser attack.
\newblock \url{https://www.owasp.org/index.php/Man-in-the-browser_attack}.

\bibitem{parno06fc}
{\sc Parno, B., Kuo, C., and Perrig, A.}
\newblock Phoolproof phishing prevention.
\newblock In {\em 10th International Conference on Financial Cryptography and
  Data Security\/} (2006), FC '06.

\bibitem{petsas15eurosec}
{\sc Petsas, T., Tsirantonakis, G., Athanasopoulos, E., and Ioannidis, S.}
\newblock Two-factor authentication: {I}s the world ready?: {Q}uantifying {2FA}
  adoption.
\newblock In {\em 8th European Workshop on System Security\/} (2015), EuroSec
  '15.

\bibitem{raguram11ccs}
{\sc Raguram, R., White, A.~M., Goswami, D., Monrose, F., and Frahm, J.}
\newblock {iSpy}: Automatic reconstruction of typed input from compromising
  reflections.
\newblock In {\em ACM Conference on Computer and Communications Security\/}
  (2011), CCS '11.

\bibitem{valiente14audiology}
{\sc Rodríguez~Valiente, A., Trinidad, A., García~Berrocal, J.~R., Górriz,
  C., and Ramírez~Camacho, R.}
\newblock Extended high-frequency (9–20 khz) audiometry reference thresholds
  in 645 healthy subjects.
\newblock {\em International Journal of Audiology 53}, 8 (2014), 531--545.

\bibitem{russell98amjphys}
{\sc Russell, D.~A., Titlow, J.~P., and Bemmen, Y.-J.}
\newblock Acoustic monopoles, dipoles, and quadrupoles: An experiment
  revisited.
\newblock {\em American Journal of Physics 67}, 8 (1999), 660--664.

\bibitem{schurmannS13tmc}
{\sc Sch{\"{u}}rmann, D., and Sigg, S.}
\newblock Secure communication based on ambient audio.
\newblock {\em {IEEE} Trans. Mob. Comput. 12}, 2 (2013), 358--370.

\bibitem{shirvanian14}
{\sc Shirvanian, M., Jarecki, S., Saxena, N., and Nathan, N.}
\newblock Two-factor authentication resilient to server compromise using
  mix-bandwidth devices.
\newblock In {\em The Network and Distributed System Security Symposium\/}
  (2014), NDSS '14.

\bibitem{shrestha14}
{\sc Shrestha, B., Saxena, N., Truong, H., and Asokan, N.}
\newblock Drone to the rescue: Relay-resilient authentication using ambient
  multi-sensing.
\newblock In {\em Financial Cryptography and Data Security\/} (2014), FC '14.

\bibitem{statcounter}
{\sc {StatCounter}}.
\newblock {StatCounter} global stats.
\newblock \url{http://gs.statcounter.com/}.

\bibitem{ansi}
{\sc {The American National Standards Insitute}}.
\newblock {ANSI} s1.11-2004 - {S}pecification for octave-band and
  fractional-octave-band analog and digital filters, 2004.

\bibitem{cherrypy}
{\sc {The CherryPy team}}.
\newblock {CherryPy}.
\newblock \url{http://www.cherrypy.org/}.

\bibitem{truong14percom}
{\sc Truong, H. T.~T., Gao, X., Shrestha, B., Saxena, N., Asokan, N., and
  Nurmi, P.}
\newblock Comparing and fusing different sensor modalities for relay attack
  resistance in zero-interaction authentication.
\newblock In {\em International Conference on Pervasive Computing and
  Communications\/} (2014), PerCom '14.

\bibitem{book}
{\sc V\'{e}r, I., and Beranek, L.}
\newblock {\em Noise and Vibration Control Engineering}.
\newblock Wiley, 2005.

\bibitem{Wang06cacm}
{\sc Wang, A.}
\newblock The shazam music recognition service.
\newblock {\em Commun. {ACM} 49}, 8 (2006), 44--48.

\bibitem{w3cwebbluetooth}
{\sc {Web Bluetooth Community Group}}.
\newblock {Web Bluetooth}.
\newblock \url{https://webbluetoothcg.github.io/web-bluetooth/}.

\bibitem{weir09compsec}
{\sc Weir, C.~S., Douglas, G., Carruthers, M., and Jack, M.~A.}
\newblock User perceptions of security, convenience and usability for ebanking
  authentication tokens.
\newblock {\em Computers {\&} Security 28}, 1-2 (2009), 47--62.

\bibitem{weir10int}
{\sc Weir, C.~S., Douglas, G., Richardson, T., and Jack, M.~A.}
\newblock Usable security: User preferences for authentication methods in
  ebanking and the effects of experience.
\newblock {\em Interacting with Computers 22}, 3 (2010), 153--164.

\bibitem{aircable}
{\sc {Wireless Cables Inc.}}
\newblock {AIRCable}.
\newblock \url{https://www.aircable.net/extend.php}.

\bibitem{yubico}
{\sc {Yubico}}.
\newblock {Yubikey hardware}.
\newblock \url{https://www.yubico.com/}.

\end{thebibliography}
}

\section*{Appendix}

\appendix

\section{System Usability Scale}
\label{app:sus}
We report the items of the System Usability Scale~\cite{sus}.
All items were answered with a 5-point Likert-scale from \emph{Strongly Disagree} to \emph{Strongly Agree.}

\begin{itemize}[noitemsep]
\small
\item[Q1] I think that I would like to use this system frequently.	
\item[Q2] I found the system unnecessarily complex.
\item[Q3] I thought the system was easy to use.
\item[Q4] I think that I would need the support of a technical person to be able to use this system.	
\item[Q5] I found the various functions in this system were well integrated.
\item[Q6] I thought there was too much inconsistency in this system.
\item[Q7] I would imagine that most people would learn to use this system very quickly.
\item[Q8] I found the system very cumbersome to use.
\item[Q9] I felt very confident using the system.
\item[Q10] I needed to learn a lot of things before I could get going with this system.
\end{itemize}

\section{Post-test Questionnaire}
\label{app:posttest}
We report the items of the post-test questionnaire.
All items were answered with a 5-point Likert-scale from \emph{Strongly Disagree} to \emph{Strongly Agree.}

\begin{itemize}[noitemsep]
\small
\item[Q1]	I thought the audio-based method was quick.
\item[Q2]	I thought the code-based method was quick.
\item[Q3]	If Second-Factor Authentication were mandatory, I would use the audio-based method to log in.
\item[Q4]	If Second-Factor Authentication were mandatory, I would use the code-based method to log in.
\item[Q5]	If Second-Factor Authentication were optional, I would use the audio-based method to log in.
\item[Q6]	If Second-Factor Authentication were optional, I would use the code-based method to log in.
\item[Q7]	I would feel comfortable using the audio-based method at home.
\item[Q8]	I would feel comfortable using the audio-based method at my workplace.
\item[Q9]	I would feel comfortable using the audio-based method in a cafe.
\item[Q10]	I would feel comfortable using the audio-based method in a library.
\item[Q11]	I would feel comfortable using the code-based method at home.
\item[Q12]	I would feel comfortable using the code-based method at my workplace.
\item[Q13]	I would feel comfortable using the code-based method in a cafe.
\item[Q14]	I would feel comfortable using the code-based method in a library.
\end{itemize}

\section{User Comments}
\label{app:comments}
This section lists some of the comments that participants added to their post-test questionnaire.

\renewenvironment{quote}{%
  \list{}{%
    \leftmargin0.2cm   
    \rightmargin\leftmargin
  }
  \item\relax
}

{\small
\begin{quote}
``Sound-Proof is faster and automatic. Increased security without having to do more things''
\end{quote}


\begin{quote}
``I would use Sound-Proof, because it is less complicated and faster.
I do not need to unlock the phone and open the application.
In a public place it would feel a bit awkward unless it becomes widespread.
Anyway, I am already logged in most websites that I use.''
\end{quote}


\begin{quote}
``I like the audio idea, because what I hate the most about second-factor authentication is to have to take my phone out or find it around.''
\end{quote}


\begin{quote}
``Sound-Proof is much easier. I am security-conscious and already use 2FA. I would be willing to switch to the audio-based method.''
\end{quote}


\begin{quote}
``I already use Google 2SV and prefer it because I think it's more secure. However, Sound-Proof is seamless.''
\end{quote}
}


\end{document}